\documentclass[twocolumn]{aastex63}
\usepackage{natbib, color}
\usepackage[normalem]{ulem}
\usepackage{epstopdf}
\usepackage{graphicx}
\usepackage{appendix}

\definecolor{red}{rgb}{1.0,0.0,0.0}

\newcommand{\Mj}[1]{$M_\mathrm{Jup}$}

\usepackage[utf8]{inputenc}

\begin{document}

\title{CD-27 11535: Evidence for a Triple System in the $\beta$ Pictoris Moving Group}

\author[0000-0002-0154-5809]{Andrew D. Thomas}
\affiliation{Kavli Institute for Particle Astrophysics and Cosmology, Stanford University, Stanford, CA 94305, USA}
\affiliation{Seattle Sounders FC, 800 Occidental Ave S, Seattle, WA 98134, USA}

\author[0000-0001-6975-9056]{Eric L. Nielsen}
\affiliation{Kavli Institute for Particle Astrophysics and Cosmology, Stanford University, Stanford, CA 94305, USA}
\affiliation{Department of Astronomy, New Mexico State University, P.O. Box 30001, MSC 4500, Las Cruces, NM 88003, USA}

\author[0000-0002-4918-0247]{Robert J. De Rosa}
\affiliation{European Southern Observatory, Alonso de C\'{o}rdova 3107, Vitacura, Santiago, Chile}

\author[0000-0003-2461-6881]{Anne E. Peck}
\affiliation{Department of Astronomy, New Mexico State University, P.O. Box 30001, MSC 4500, Las Cruces, NM 88003, USA}

\author[0000-0003-1212-7538]{Bruce Macintosh}
\affiliation{Kavli Institute for Particle Astrophysics and Cosmology, Stanford University, Stanford, CA 94305, USA}
\affiliation{University of California Observatories, 1156 High Street, Santa Cruz, CA 95064, USA}
\affiliation{Department of Astronomy and Astrophysics, UC Santa Cruz, Santa Cruz CA 95064}

\author[0000-0001-6305-7272]{Jeffrey Chilcote}
\affiliation{Department of Physics and Astronomy, University of Notre Dame, 225 Nieuwland Science Hall, Notre Dame, IN, 46556, USA}

\author[0000-0002-6221-5360]{Paul Kalas}
\affiliation{Department of Astronomy, University of California, Berkeley, CA 94720, USA}

\author[0000-0003-0774-6502]{Jason J. Wang}
\affiliation{Center for Interdisciplinary Exploration and Research in Astrophysics (CIERA), Northwestern University, Evanston, IL 60208, USA}
\affiliation{Department of Physics and Astronomy, Northwestern University, Evanston, IL 60208, USA}
\affiliation{Department of Astronomy, California Institute of Technology, Pasadena, CA, 91125, USA}

\author[0000-0002-3199-2888]{Sarah Blunt}
\affiliation{Center for Interdisciplinary Exploration and Research in Astrophysics (CIERA), Northwestern University, Evanston, IL 60208, USA}
\affiliation{Cahill Center for Astronomy $\&$ Astrophysics, California Institute of Technology, Pasadena, CA 91125, USA}

\author[0000-0002-7162-8036]{Alexandra Greenbaum}
\affiliation{IPAC, Mail Code 100-22, Caltech, 1200 E. California Blvd, Pasadena, CA 91125, USA}

\author[0000-0002-9936-6285]{Quinn M. Konopacky}
\affiliation{Center for Astrophysics and Space Sciences, University of California, San Diego, La Jolla, CA 92093, USA}

\author[0000-0002-6194-043X]{Michael J. Ireland}
\affiliation{Research School of Astronomy $\&$ Astrophysics, Australian National University, Canberra, ACT 2611, Australia}

\author[0000-0001-7026-6291]{Peter Tuthill}
\affiliation{Sydney Institute for Astronomy, University of Sydney, Physics Rd., NSW 2006, Australia}

\author[0000-0002-4479-8291]{Kimberly Ward-Duong}
\affiliation{Department of Astronomy, Smith College, Northampton MA 01063 USA}

\author[0000-0001-8058-7443]{Lea A. Hirsch}
\affiliation{Department of Chemical $\&$ Physical Sciences, University of Toronto Mississauga, Mississauga, ON L5L 1C6, Canada}

\author[0000-0002-1483-8811]{Ian Czekala}
\altaffiliation{NASA Hubble Fellowship Program Sagan Fellow}
\affiliation{Department of Astronomy and Astrophysics, 525 Davey Laboratory, The Pennsylvania State University, University Park, PA 16802, USA}
\affiliation{Center for Exoplanets and Habitable Worlds, 525 Davey Laboratory, The Pennsylvania State University, University Park, PA 16802, USA}
\affiliation{Center for Astrostatistics, 525 Davey Laboratory, The Pennsylvania State University, University Park, PA 16802, USA}
\affiliation{Institute for Computational \& Data Sciences, The Pennsylvania State University, University Park, PA 16802, USA}
\affiliation{Department of Astronomy, 501 Campbell Hall, University of California, Berkeley, CA 94720-3411, USA}

\author[0000-0001-7016-7277]{Franck Marchis}
\affiliation{SETI Institute, 339 Bernardo Ave, Suite 200, Mountain View, CA 94043, USA}

\author[0000-0002-4164-4182]{Christian Marois}
\affiliation{National Research Council of Canada, Herzberg Astronomy and Astrophysics, 5071 West Saanich Rd, Victoria, BC, V9E 2E7, Canada}
\affiliation{University of Victoria, 3800 Finnerty Rd, Victoria, BC, Canada V8P 5C2}

\author[0000-0001-6205-9233]{Max A. Millar-Blanchaer}
\affiliation{Department of Physics, University of California, Santa Barbara, CA 93106, USA}

\author[0009-0008-9687-1877]{William Roberson}
\affiliation{Kavli Institute for Particle Astrophysics and Cosmology, Stanford University, Stanford, CA 94305, USA}

\author[0000-0002-9156-9651]{Adam Smith}
\affiliation{Department of Astronomy, New Mexico State University, P.O. Box 30001, MSC 4500, Las Cruces, NM 88003, USA}

\author[0009-0000-8603-169X]{Hannah Gallamore}
\affiliation{Department of Astronomy, New Mexico State University, P.O. Box 30001, MSC 4500, Las Cruces, NM 88003, USA}

\author[0000-0003-3906-9518]{Jessica Klusmeyer}
\affiliation{Department of Astronomy, New Mexico State University, P.O. Box 30001, MSC 4500, Las Cruces, NM 88003, USA}

\begin{abstract}
We present new spatially resolved astrometry and photometry of the CD-27 11535 system, a member of the $\beta$ Pictoris moving group consisting of two resolved K-type stars on a $\sim$20-year orbit. We fit an orbit to relative astrometry measured from NIRC2, GPI, and archival NaCo images, in addition to literature measurements. However, the total mass inferred from this orbit is significantly discrepant from that inferred from stellar evolutionary models using the luminosity of the two stars. We explore two hypotheses that could explain this discrepant mass sum; a discrepant parallax measurement from {\it Gaia} due to variability, and the presence of an additional unresolved companion to one of the two components. We find that the $\sim$20-year orbit could not bias the parallax measurement, but that variability of the components could produce a large amplitude astrometric motion, an effect which cannot be quantified exactly without the individual {\it Gaia} measurements. The discrepancy could also be explained by an additional star in the system. We jointly fit the astrometric and photometric measurements of the system to test different binary and triple architectures for the system. Depending on the set of evolutionary models used, we find an improved goodness of fit for a triple system architecture that includes a low-mass ($M=0.177\pm0.055$\,$M_{\odot}$) companion to the primary star. Further studies of this system will be required in order to resolve this discrepancy, either by refining the parallax measurement with a more complex treatment of variability-induced astrometric motion, or by detecting a third companion.
\end{abstract}

\section{Introduction}
Knowing the precise age of stars is essential to understanding the planets they host. The age of a star, especially a star evolving toward the zero-age main sequence, can be found using evolutionary models based on its mass and absolute luminosity, if the mass is measured independently. As the stars in moving groups formed at the same time, the age of one star provides an estimate of the age of the whole group, especially when the moving group is young (\cite{Zuckerman:2004}, \cite{bell:2015}, \cite{schlieder:2016}). Exoplanet formation and brown dwarf evolution is closely related to that of its host star. Hence, knowing the age of a moving group can reveal the age of various exoplanets and brown dwarfs in it. These accurate ages allow us to constrain evolutionary models for exoplanet and brown dwarfs \citep{Burgasser2009} or simply lead to a better understanding of the formation of specific substellar objects. Binary stars provide the opportunity to discover the mass of the system through direct observation of the orbit. Fitting for the orbit reveals the dynamical total mass which, in turn, reveals the age \citep{Nielsen2016}. 

CD-27 11535 (CD-27) is a young K5 star and member of the $\beta$ Pictoris ($\beta$~Pic) moving group \citep{Torres:2006}, with a \textit{Gaia} Data Release 2 (DR2) distance of $\sim83.3$pc \citep{GAIADATADR2}. CD-27 was first identified as a young star by \cite{Torres:2006} as part of the Search for Associations Containing Young stars (SACY) database, the aim of which was to determine space motion of young stars through radial velocity, derived from spectroscopic observations. The spectroscopic observations also allowed them to determine the spectral type and equivalent width of Li 6708$\text{\r{A}}$, which can be used to suggest the age of the star \citep{Carlos:2016}. For CD-27, they report an equivalent width of 490m$\text{\r{A}}$ for Lithium - certainly indicating youth. 

\begin{deluxetable}{lccc}
\tabletypesize{\footnotesize}
\tablecaption{Parameters of CD-27 11535}
\label{tab:basicparameters}
\tablewidth{0pt}
\tablehead{\colhead{Parameter} & \colhead{Value} & \colhead{Ref.}}
\startdata
Alternate names & TYC 6820-223-1 & \\
& Gaia DR2/3 4107812485571331328 & \\
& WDS 17151-2750 & \\
& ELP 40 & \\
Position & & a \\
\quad\quad Right Ascension & 17:15:$03.61 \pm 0.26$\,mas & \\
\quad\quad Declination & $-27$:49:$39.74 \pm 0.19$\,mas & \\
Proper Motion (mas\,yr$^{-1}$) & $4.531 \pm 0.447$, $-45.822 \pm 0.297$ & a\\
Parallax (mas) & $12.0040 \pm 0.2854$ & a \\
Apparent magnitudes & & \\
\quad\quad $B$ & $11.68 \pm 0.12$ & b \\
\quad\quad $V$ & $10.596 \pm 0.063$ & c \\
\quad\quad $G$ & $10.091995 \pm 0.004247$ & d \\
\quad\quad $I$ & $9.129 \pm 0.032$ & c \\
\quad\quad $J$ & $8.174 \pm 0.029$ & e \\
\quad\quad $H$ & $7.537 \pm 0.033$ & e \\
\quad\quad $K_s$ & $7.384 \pm 0.031$ & e \\
Spectral type & K5Ve & \\
\enddata
\tablenotetext{a}{\cite{GAIADATADR2}}
\tablenotetext{b}{\citet{Hog:2000}} 
\tablenotetext{c}{\citet{Kiraga:2012}}
\tablenotetext{d}{\citet{GAIAEDR3:2021}}
\tablenotetext{e}{\citet{Cutri:2003}}
\end{deluxetable}

Age estimates of the CD-27 system of 6\,Myr by \cite{Weise:2010} and 10\,Myr by \cite{elliott:2014} both reference the lithium equivalent width measurement by \cite{Torres:2006}. While an equivalent width of 490\,m\AA~does indicate an extremely young age \citep{Stanford-Moore:2020}, a confounding factor is stellar multiplicity which can affect the interpretation of such a measurement. Following its identification as a young star by \cite{Torres:2006}, \cite{Song:2012} obtained new spectra, and associated the star with the Upper Scorpius subgroup of the Scorpius-Centaurus (Sco-Cen) complex, though they note that the star may be a nearby non-member. Later, \cite{elliott:2014} use radial velocity (RV) measurements to approximate the systemic velocity and employ the convergence method of kinematic trace-back by \cite{Torres:2006} to determine group membership. They conclude that CD-27 is a candidate member of $\beta$~Pic, a doubt which stems from the uncertainty in the systemic velocity without an orbital solution. This is further studied by \cite{Shkolnik:2017} who assign $\beta$~Pic group membership by their confirmation criteria based on spectral type, proper motion, Li and H$_\alpha$ equivalent widths and RV measurements. Age estimates of the $\beta$~Pic moving group by \cite{Nielsen2016} suggest an age of $26\pm3$\,Myr, as such we expect that of CD-27 to be similar.

CD-27 was first identified as a binary by \cite{elliott:2014} who noted a significant RV difference measured by \cite{Torres:2006} and \cite{Song:2012} ($-6.4 \pm 1.0$ and $-1.1 \pm 1.8$\,km\,s$^{-1}$, respectively), suggesting the system is a single-lined spectroscopic binary. Their analysis yielded an RV of $-6.9 \pm 1.4$\,km\,s$^{-1}$, closer to the value reported by \cite{Torres:2006} but significantly different than the $-12.3 \pm 3.2$\,km\,s$^{-1}$ reported by \textit{Gaia} Data Release 2 (DR2; \citealp{GAIADATADR2}). \cite{Elliot:2015} subsequently resolved the system with adaptive optics (AO) observations with an angular separation of $0\farcs08$, listing the system as “AB, SB1.” \cite{Alonso:2015}, citing \cite{Elliot:2015}, list the system as an AaAb-B triple system, though it is unclear if this is based on additional data not in either paper, or a misreading of the “AB, SB1" notation. More recently, \cite{Bonavita:2022} as part of the SpHere INfrared Exoplanet (SHINE) project also studied the CD-27 system. They fit an orbit to their relative astrometric measurements, yielding a total mass of the system of $2.14\pm0.27$\,$M_\odot$, $\sim1.5\sigma$ larger than their expectation for the total mass of the two stars of $1.75$--$1.82$\,$M_\odot$ from evolutionary models, depending on the age. 

Our investigation into the CD-27 system was initially intended to provide another benchmark system for the $\beta$~Pic moving group, utilizing the method outlined by \cite{Nielsen2016} and \cite{montet:2015} to further constrain the age the association. However, the significant discrepancy between the total mass from the orbit and the masses derived from the luminosity of each component and evolutionary models precluded this analysis. Instead, we explored two scenarios that could explain this discrepancy. We begin with a description of the high angular resolution imaging observations of the system in ~\ref{Obs and Data Red}, which we combined with literature measurements to fit an orbit to the relative astrometry in Section~\ref{Orbit Fitting}. We consider the possibility of an additional component within the system and the possibility of an incorrect distance determination in Section~\ref{sec:scenarios}. We derive limits on the presence of an additional companion in Section~\ref{sec:limits}, and give conclusions and possible next steps for further studies in Section~\ref{sec:conclusion}.

\section{Observations and Data Reduction}
\label{Obs and Data Red}
\begin{figure*}
\centering
\includegraphics[clip,width=1.0\textwidth]{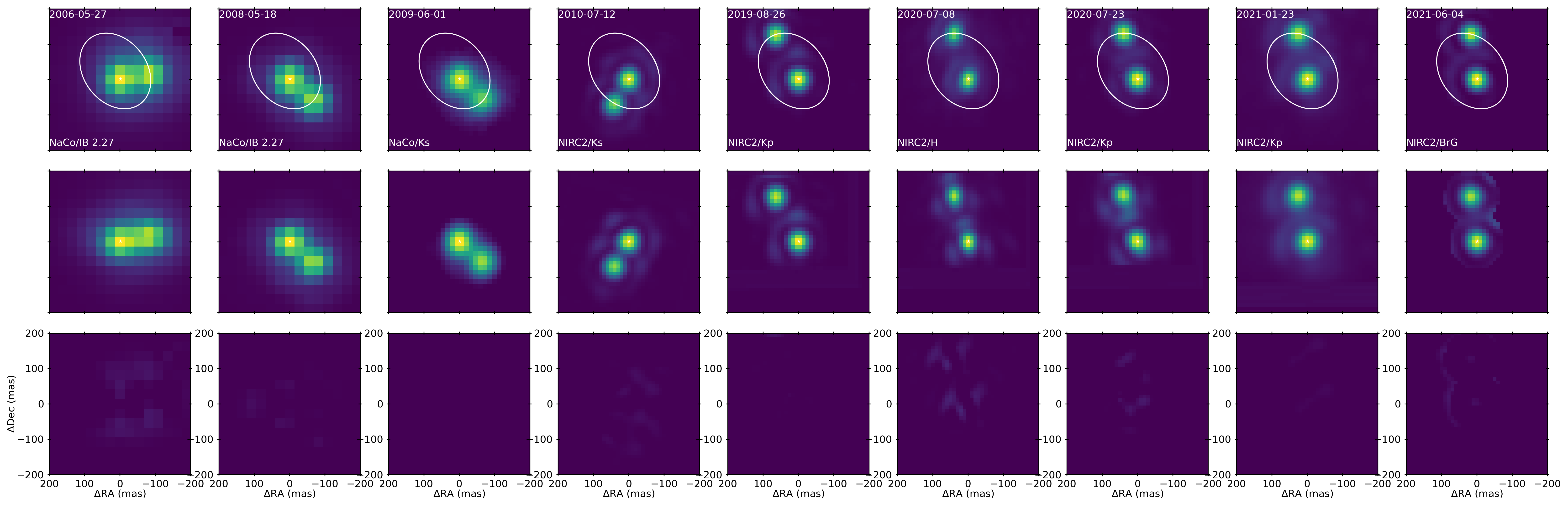}
\caption{CD-27 11535 for our nine epochs of imaging data, excluding the two GPI epochs. Along the top row are the initial reduced images, the middle row displays the model created using a PSF star while the bottom row shows the results of the PSF fitting and subtraction of the model. The best fit orbit (white) is displayed on the reduced images. The labels at the top of each column indicate the epoch while those at the bottom of the reduced image list the instrument and filter.}
\label{fig:gallery}
\end{figure*}

To constrain the visual orbit of CD-27 we obtained diffraction-limited images of the system with the Keck/NIRC2 (PI: K. Matthews) and Gemini-S/GPI \citep{Macintosh:2014js} instruments. We complemented these observations with archival datasets from VLT/NaCo \citep{Lenzen:2003iu,Rousset:2003hh}. In total, we collected 18 observations on 11 separate epochs. The NaCo and NIRC2 observations were obtained in a conventional imaging mode, whereas the GPI observations were obtained with a non-redundant aperture mask \citep{Greenbaum:2019hr}. A summary of these imaging observations are given in Table~\ref{tab:observing_log}. We also obtained high-resolution optical echelle spectra of the system with APO/ARCES \citep{Wang:2003} as a part of a larger radial velocity monitoring program.

\subsection{VLT/NaCo Imaging}
Observations of CD-27 that were made between 2006 May 27 and 2009 Jun 1 with the NaCo instrument were obtained from the ESO Science Archive Facility\footnote{\url{http://archive.eso.org}}. Two datasets were obtained with the S27 camera (27\,mas/px) and the IB2.27 intermediate-band filter, and one was obtained with the S13 camera (13\,mas/px) and the $K_s$ broad-band filter. These data were obtained under program IDs 077.C-0483, 081.C-0825, and 083.C-0659. The first two epochs were originally published in \cite{Elliot:2015}, however the large uncertainty of their separation measurements motivated us to re-analyze their observations.

We used a standard near-infrared data reduction process to reduce and analyse each dataset using associated calibrations obtained from the archive. The dark current was measured by finding the median of several exposures of the same exposure time taken while the instrument light path was blocked. This combined frame was then subtracted from each raw image. We created a flat field and bad pixel map using pairs of lamp-on and lamp-off exposures obtained during the day. These were used to perform the flat field correction of the raw images, and to correct for bad pixels. We did not apply any distortion correction to the image as the effect of distortion over such a small angular separation between the two resolved components of the CD-27 system is negligible. The thermal background within each image was estimated using a median combination of all images, which was then subtracted from each image.

Relative astrometric and photometric measurements were made by fitting the point spread function (PSF) of both components with that of a single star observed on the same (or neighbouring) night with the same instrument configuration. We searched the archive for suitable observations for each of our three datasets. Unlike for CD-27, the individual images for the reference stars were aligned and combined to create a high signal-to-noise (SNR) PSF that could be used for PSF fitting. The centroid of the star in each image was calculated by fitting a two-dimensional Gaussian within a region of four pixels centered on an initial guess for the location of the star. This process was repeated using the result of the first iteration as the guess position for the second.

A model of the two components of the CD-27 system was constructed using two copies of the model PSF interpolated to the positions $(x_0, y_0)$ and $(x_1, y_1)$, and multiplied by scaling factors $f_0$ and $f_1$. A background term $(A)$ was added to account for any residual background signal present in the image. We used the implementation of Powell's method within the \texttt{scipy.optimize} package to find the optimal set of parameters that minimized the residuals after model subtraction in two $1.5\lambda/D$ apertures centered on the positions of the two components. This was repeated for each of the potential PSF calibrator stars. The PSF star that resulted in the smallest amplitude residuals was used to measure the relative astrometry and photometry between the resolved components.

The pixel positions of the two components were converted into an angular separation and position angle using the plate scale and orientation of the detector given in Table~\ref{tab:astrometry}. Magnitude differences were calculated from the ratio of the two scaling factors. This process was repeated for each image of the CD-27 system, yielding an average and standard deviation for each value that are given in Table~\ref{tab:astrometry}. The reduced images, the best-fit PSF model, and the corresponding residuals are shown in Figure~\ref{fig:gallery}.

\subsection{Keck/NIRC2 Imaging}
The system was observed on six separate epochs over the course of eleven years with Keck/NIRC2 through seven different filters ($z$, $J$, $H$, $K_s$, $K'$, $L'$, $BrG$). A single star was observed with the same filter that could act as a PSF reference in all but the last epoch in June 2021. For that epoch, a PSF was constructed using the binary 1RXS J195602.8-320720 after digitally masking the fainter secondary. The observations were reduced using a standard near-infrared reduction pipeline similar to that employed to reduce the NaCo data. The raw images were dark subtracted, divided by a flat field, and cleaned of bad pixels using outlier rejection. The geometric distortion of the detector was corrected using a flux-conserving algorithm \citep{Yelda2011, Service2016}. The relative position and brightness of the two resolved components of the CD-27 system were calculated using the same PSF fitting approach used for both the NaCo and GPI datasets. This process was repeated for each image yielding an average value and standard deviation for the pixel offsets and flux ratio. These were converted into on-sky separations and position angles using the calibration values listed in Table~\ref{tab:astrometry}. The reduced images, the best-fit PSF model, and the corresponding residuals for a subset of the observations of the system are shown in Figure~\ref{fig:gallery}.

\subsection{Gemini-S/GPI Aperture Masking}
\begin{figure}
\centering
\includegraphics[width=\columnwidth]{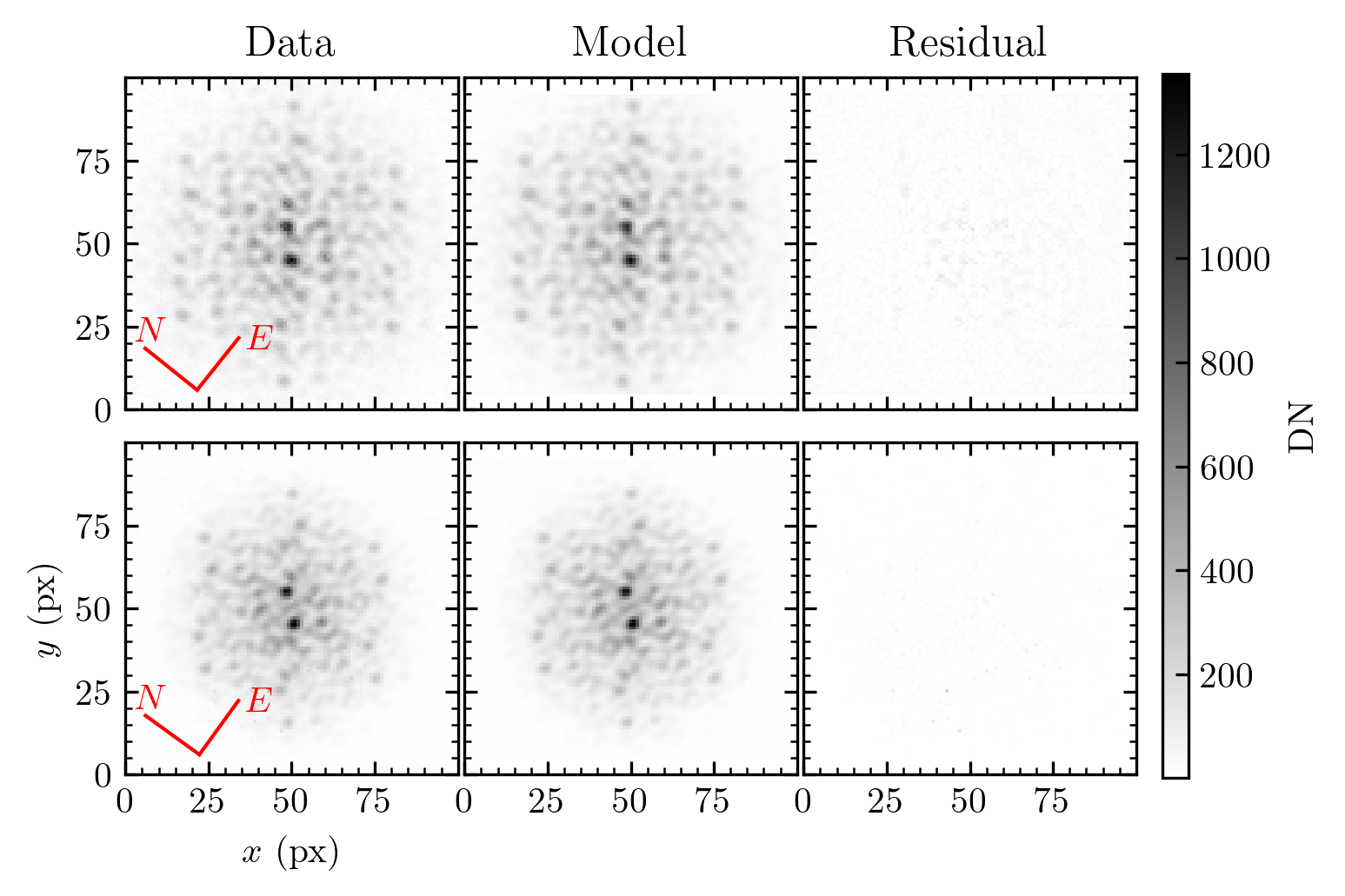}
\caption{GPI NRM images of CD-27 11535 (left), a model constructed from the calibrator star HD 153318 (middle), and the residual of the two (right) for the 2018 Aug 15 (top) and 2019 Aug 05 (bottom) epochs.}
\label{fig:gpi-nrm}
\end{figure}
CD-27 was observed with GPI on 2018 August 15 and 2019 August 05 alongside the PSF calibrator HD 153318 (see Table~\ref{tab:observing_log}). The GPI images were all processed and reduced with the GPI Data Reduction Pipeline (DRP) v1.5.0 \citep{Perrin:2014}. After dark current subtraction and bad pixel interpolation, microspectra within the raw 2-D image were extracted to create an ($x$, $y$, $\lambda$) data cube \citep{Maire:2014gs}. An image of an argon arc lamp was taken immediately before the data to calibrate the wavelength axis of the data cube. Bad pixels were corrected using outlier rejection both before and after the data cube construction step. The system was clearly resolved into two components within the reduced image (Fig.~\ref{fig:gpi-nrm}), at an angular separation far beyond what is typically explored with the non-redundant mask (NRM) technique. These observations were scheduled prior to a reliable determination of the orbit of the system.

Rather than using a specialized pipeline to reduce these observations (e.g. \citealp{Greenbaum:2019hr}), we instead used a similar approach as for the NaCo observations described in the previous section because of the large angular separation between the two components (Fig~\ref{fig:gpi-nrm}). Observations of the single calibrator star HD 153318 were reduced and used to construct a model of the PSF. Two copies of this PSF were shifted and scaled to fit the two resolved components of the CD-27 system. This process was repeated using the central wavelength slice of each observation from each epoch (19 in 2018, 16 in 2019) yielding an average and standard deviation for each measurement. On-sky separations and position angles were calculated using the relevant calibration measurements given in Table~\ref{tab:astrometry}. The reduced images, the best-fit PSF model, and the corresponding residuals are shown in Figure~\ref{fig:gpi-nrm}.

\subsection{APO/ARCES Spectra}
We obtained optical echelle spectra of CD-27 at the Apache Point Observatory (APO) 3.5m telescope with the ARCES instrument \citep{Wang:2003}, on five epochs in 2021. Given the seeing-limited conditions and the close separation of the system, we obtained a blended spectrum of the entire system. The observations were reduced using a standard spectroscopic reduction pipeline that performs bias and flat-field correction, measures and corrects the continuum, and derives a wavelength solution from observations of a ThAr calibration lamp.

\begin{figure}
\includegraphics[width=\columnwidth]{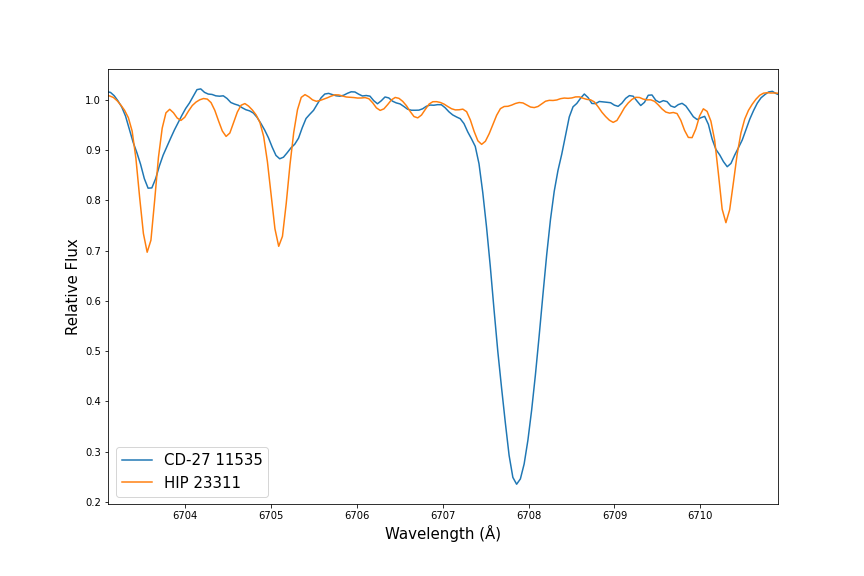}
\caption{APO/ARCES spectra of CD-27 11535 from 2021 June 20, showing the deep 6707.79$\text{\r{A}}$ lithium line.  Also plotted for comparison is the older K3 star HIP 23311, which has no significant lithium absorption.  While the CD-27 spectrum is a composite of multiple stars of varying spectral types on the slit, the strong lithium absorption is compelling evidence for a young ($\lesssim$100 Myr) age for the system.}
\label{fig:lithium_line}
\end{figure}
Figure~\ref{fig:lithium_line} shows the region surrounding the lithium 6707.79\,\AA~line for the CD-27 system taken on 2021 June 20, and for our RV standard HIP 23311 with a spectral type of K3 taken on the same night. Despite the near-equal visual magnitude of A and B, we do not observe double lines in our spectrum, but rather broad lines suggesting some or all of the components are rapid rotators. No significant RV shift is measured over our four months of data. We confirm the strong lithium absorption observed by \citet{Torres:2006}, though with these unresolved spectra we cannot determine how much lithium corresponds to each component of the system.  Nevertheless, given the late spectral types of the components, with a K5 spectral type assigned to the system, this strong lithium absorption is a clear sign of youth, consistent with $\beta$ Pictoris moving group membership.

\section{Visual Orbit of CD-27 11535}
\label{Orbit Fitting}
The relative astrometry derived from our analysis and the literature astrometric measurements are compiled in Table~\ref{tab:astrometry}. The speckle interferometry measurements from SOAR span 2016 through 2023, and we applied a correction of quadrant for consistency with our measurements.

\subsection{Orbit Fitting}
With the astrometry for each epoch collected, we proceed with fitting a Keplarian orbit to the relative astrometry of B with respect to A. To do so, we use a parallel-tempered Markov Chain Monte Carlo (MCMC) procedure \citep{Foreman-Mackey2013}, implemented within the orbit fitting software $\texttt{orbitize!}$ \citep{Blunt:2017}. In total, 256 chains are run in parallel at 16 temperatures to sample the posterior distribution of the following eight parameters: semi-major axis ($a$), eccentricity ($e$), inclination angle ($i$), argument of periastron ($\omega$), position angle of nodes ($\Omega$), normalized epoch of periastron passage ($\tau$), parallax ($\pi$) and total mass ($M_{tot}$). Here, $\tau$ is the fraction of the orbit past a given reference date (in our case 2020 Jan 01), and is parameterized by the period \citep{Blunt2020}.

To ensure rapid convergence of the chains, the walkers are initialized from an optimized starting position found using a rejection sampling algorithm (OFTI) to find 256 potential orbits based on the first three NaCo epochs. For the OFTI algorithm, we employ a uniform prior in total mass between 0.5 and 5\,$M_{\odot}$. The semi-major axis prior is a log uniform distribution with a minimum value of $1\times10^{-3}$\,au and a maximum of $1\times10^7$\,au. Eccentricity and epoch of periastron are given uniform priors between 0.0 and 1.0. The argument of periastron and position angle of nodes also have uniform priors between 0.0 and $2\pi$ but the inclination angle makes use of a sine prior. Finally, we use a Gaussian prior on the parallax from the parallax and error of $12.0040\pm0.2854$\,mas (DR2; \cite{GAIADATADR2}). The priors used in the MCMC analysis were identical, other than that of the semi-major axis which had a log uniform prior between $1\times10^{-1}$\,au and $1\times10^3$\,au.

\begin{deluxetable*}{lcccc}
\tabletypesize{\normalsize}
\tablecaption{Best-fit orbital parameters and corresponding median and 1-$\sigma$ confidence intervals for the CD-27 system. Values from \cite{Bonavita:2022} given for comparison.}
\label{tab:orbit}
\tablewidth{0pt}
\tablehead{\colhead{Parameter} & \colhead{Unit} & \colhead{$\chi_{\rm min}^2$ Orbit} & \colhead{Median, 1-$\sigma$ CI} & \colhead{\cite{Bonavita:2022}}}
\startdata
$a$           & au          & 9.7 & $9.8_{-0.2}^{+0.2}$ & $9.73\pm0.15$\\
$e$           & \nodata     & 0.2396 & $0.2394_{-0.0005}^{+0.0005}$ & $0.264\pm0.026$\\
$i$           & deg         & 141.39 & $141.27_{-0.15}^{+0.15}$ & $147.2\pm5.9$\\
$\omega$      & deg         & 11.0 & $11.6_{-0.4}^{+0.5}$ & $12.8\pm8.8$\\
$\Omega$      & deg         & 218.1 & $218.4_{-0.2}^{+0.2}$ & $35.3\pm4.8$\\
$T_0$         & yr          & 2029.03 & $2029.06_{-0.03}^{+0.03}$ & $2009.35\pm0.24$\\
$P$           & yr          & 19.81 & $19.83_{-0.02}^{+0.02}$ & $20.78\pm0.84$\\
$M_{\rm tot}$ & $M_{\odot}$ & 2.35 & $2.36_{-0.16}^{+0.17}$ & $2.14\pm0.27$\\
\enddata
\end{deluxetable*}
\begin{figure}
\includegraphics[width=\columnwidth]{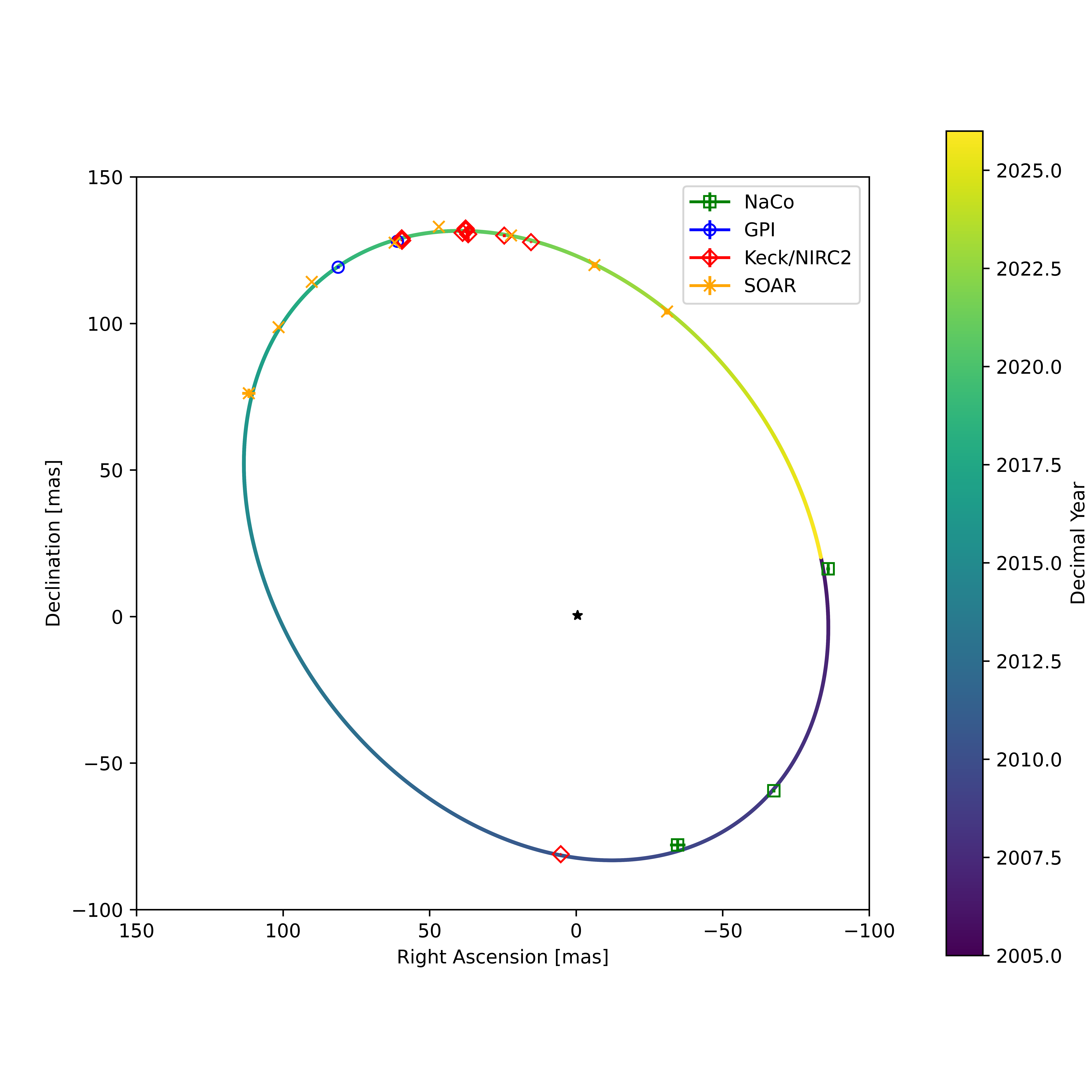}
\caption{Best-fit visual orbit of CD-27 11535 system found using \texttt{orbitize!}. Different symbols and colors show the observations taken by the various instruments while the color-bar shows the relative position of the visible components over time.}
\label{fig:SkyPlot}
\end{figure}
\begin{figure}
\label{fig:seppavstime}
\includegraphics[width=\columnwidth]{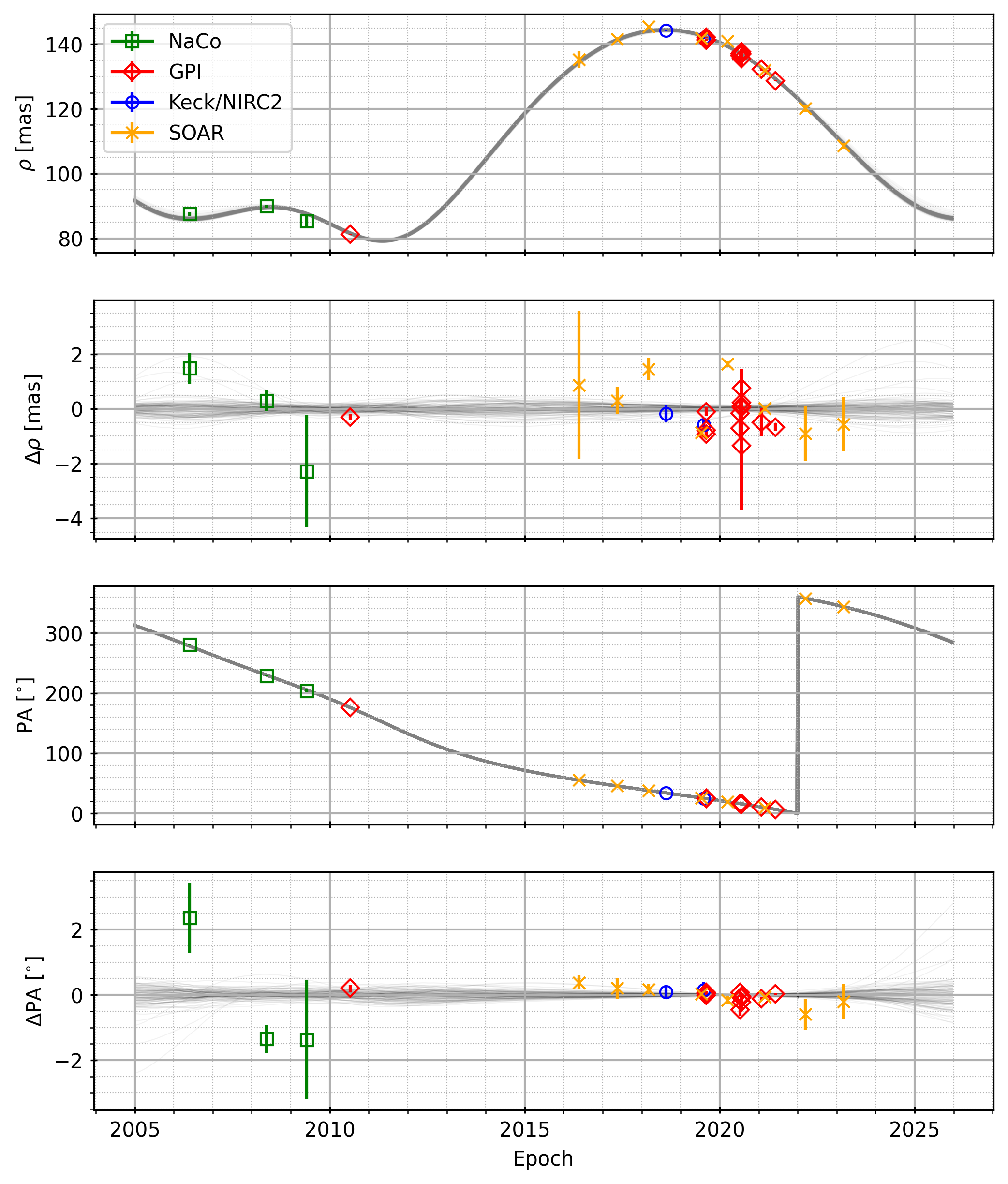}
\caption{Astrometric measurements (symbols) and a hundred orbits randomly sampled from the MCMC posterior distributions (gray curves) showing the separation (first row) and position angle (third row) as a function of time. The second and fourth rows show the corresponding residuals relative to the median orbit.}
\label{fig:OrbitSepPa}
\end{figure}

Posteriors from the MCMC fit are given in Table~\ref{tab:orbit} and the orbit is plotted in Figures~\ref{fig:SkyPlot} and \ref{fig:OrbitSepPa}. The astrometric measurements span almost a complete orbital period resulting in a very well-constrained orbit with a reduced $\chi^2$ of 9.8. We find a fairly low eccentricity $0.2394\pm 0.0005$ and an inclination of $141.27^{\circ}\pm0.15^{\circ}$. The semi-major axis of the orbit is $9.8\pm0.2$\,au, with a period of $19.83\pm0.02$\,yr giving a total mass of $2.36_{-0.16}^{+0.17}$\,$M_{\odot}$.

\subsection{Comparison with Previous Studies}
Recently, the CD-27 system was included in a sample of binaries characterized with VLT/SPHERE \citep{Bonavita:2022} who presented an initial determination of the orbit of this system. With our improved coverage of the orbit we are able to further constrain the orbital parameters. Although we find similar values for the orbital period and eccentricity (see Table~\ref{tab:orbit}), the orbit appears rotated by 180 degrees on the sky. This discrepancy is due the difference between the studies for which component in the system is designated the primary. \citet{Bonavita:2022} report an ambiguity in which of the two components is brightest in the near-IR, whereas we are able to differentiate which of the two resolved components is the brightest at a very high confidence ($\sim$100$\sigma$ in the 2019 Aug 26 epoch).

\subsection{Comparison with Evolutionary Models}
\begin{figure*}
\label{fig:extramass}
\centering
\includegraphics[scale = 0.5]{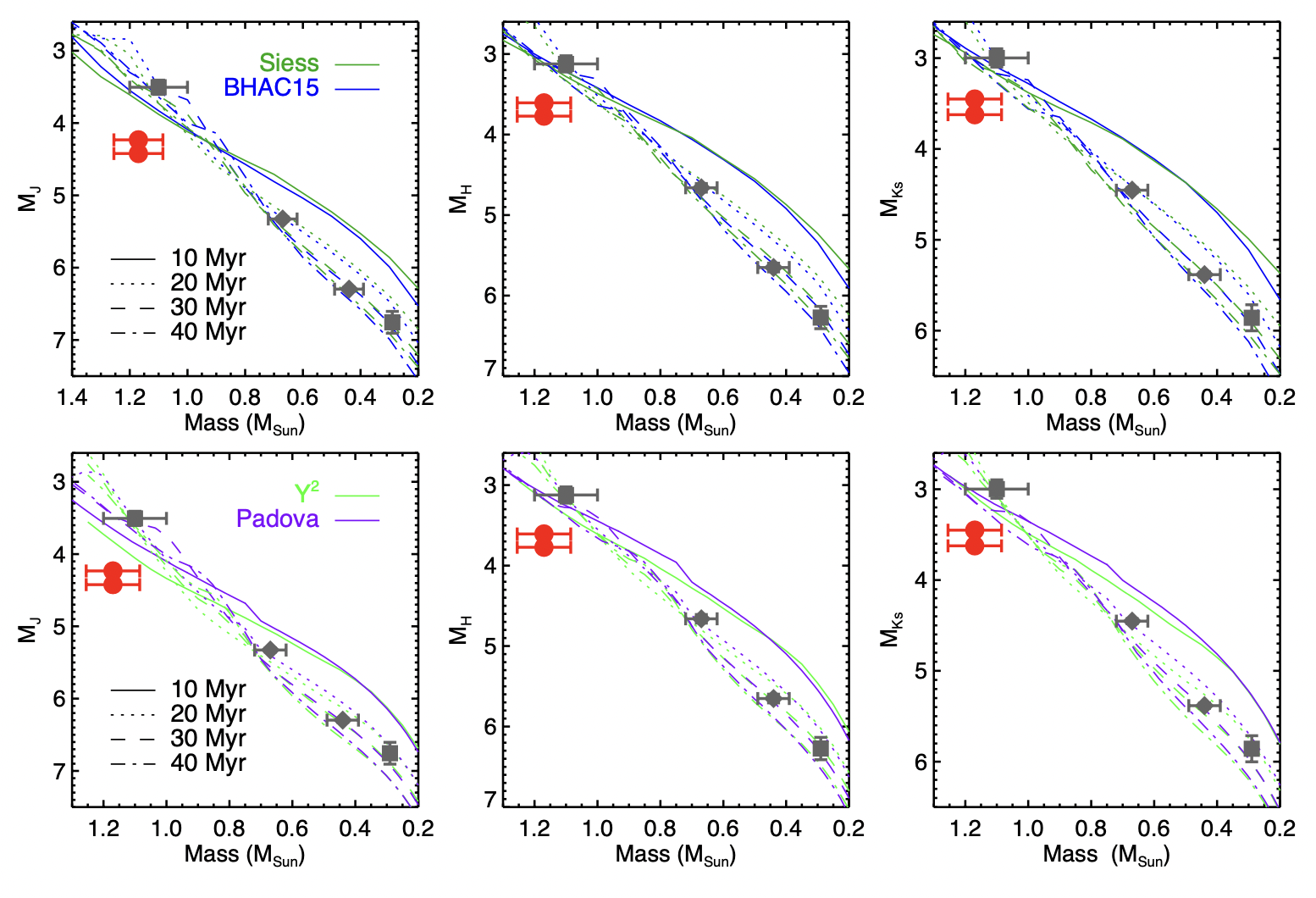}
\caption{Comparison of the measured photometry and mass of the two resolved components of the CD-27 11535 system (red circles) to four sets of theoretical models, as well as GJ 3305 A/B (gray diamonds) from \cite{montet:2015} and V343 Nor Aa/Ab (gray squares) from \cite{Nielsen2016}. For the plotted isochrones, colors corresponds to different models, and line-styles to ages.}
\label{fig:ExtraMass}
\end{figure*}
We compare the total system mass from the visual orbit to the masses predicted from four evolutionary model grids. As we do not have a measurement of the mass ratio of the system, we assume a mass ratio of $q=1$, a reasonable assumption given the near-equal flux ratio. The absolute magnitude of each component was estimated from the apparent magnitude of the blended system reported in the 2MASS catalogue \citep{Cutri:2003}, the flux ratio measured in the corresponding filter reported in Table~\ref{tab:astrometry}, and the distance derived from the {\it Gaia} DR2 parallax measurement. We generated mass-magnitude relations from four evolutionary models between 10--40\,Myr; Siess \citep{siess:2000}, BHAC15 \citep{baraffe:2015}, Yonsei-Yale \citep{spada:2013} and Padova PARSEC \citep{bressan:2012}. These models are compared to the mass and absolute magnitudes of the two components of CD-27 in Figure~\ref{fig:ExtraMass}.

Given the expected age of $26\pm3$\,Myr for the system, the two components appear either under-luminous or too massive when compared to the evolutionary models. Such a discrepancy is difficult to explain without invoking a systematic error in either---or both ---of the component masses or absolute magnitudes. If the two stars were instead above the zero-age main sequence, that could more easily be explained by a different age of the system. This discrepancy is not sensitive to our assumed mass ratio, as changing this simply shifts the components in opposite directions horizontally relative to the model tracks. This would indeed move one of the two stars closer to the main sequence but the other would become an even greater outlier.

\section{Possible Causes of the Discrepancy}
\label{sec:scenarios}
\subsection{An Additional Component}
\label{Flux Ratio Fitting}
A low-mass third companion to either of the two resolved components of the system could explain the discrepancy between the total mass from the orbit, and the mass predicted from the absolute magnitude of each component. We test this hypothesis by simultaneously fitting the visual orbit and the spectral energy distribution of both of the resolved components under various assumptions regarding the system architecture. Specifically, we investigate a binary (AB) and a triple with the additional companion around either the primary (AaAb-B), or the secondary (A-BaBb). We also test the effect of removing the \textit{Gaia} DR2 parallax constraint on the quality of the fit.
We use Metropolis-Hastings MCMC implemented within the \texttt{emcee} package \citep{Foreman-Mackey2013} to simultaneously fit the visual orbit, the flux ratios between the two resolved components, and the total near-infrared flux of the system. The visual orbit is fit as before (Section~\ref{Orbit Fitting}). The flux ratios and total flux of the system are estimated from the four aforementioned evolutionary models, with an independent fit performed per model grid. Due to the 2.9\,d variability of the system \citep{Kiraga:2012}, we only use flux ratio measurements taken on 2020 July 24 that were taken in five different bands within an hour and, consequently, would be less affected by variability.
For each fit, we run 128 chains in parallel for $10^4$ steps. The walkers are initialised with uniform distributions, centred at the values reported in Section~\ref{Orbit Fitting}. At each step in the fit, the flux of each constituent star is generated using an interpolating function given its age, mass and the evolutionary model, from which the flux ratio between the two resolved components is determined. The absolute magnitude of the blended system is calculated and converted into an apparent magnitude using the parallax, which is then compared to the 2MASS measurements for the system \cite{Cutri:2003}. We use priors to keep the age between 10 and 100 Myr (consistent with the amount of lithium absorption observed), the parallax positive, and the mass of the constituent stars between 0.1 and 1.5 $M_{\odot}$. The priors on the orbital parameters are as used in Section~\ref{Orbit Fitting}. The lower bound on the star mass prior was set by the minimum mass reported in the evolutionary models.

\begin{deluxetable*}{lllcccccc}
\label{tab:possibleconfigs}
\tabletypesize{\footnotesize}
\tablecaption{Possible configurations of CD-27 11535. $M_1$ is A or Aa, $M_2$ refers to B or Ba while $M_3$ is the mass of Ab or Bb.}
\label{tab:configresults}
\tablewidth{0pt}
\tablehead{\colhead{Configuration} & \colhead{Prior on Parallax} &  \colhead{Evolutionary Model} & \colhead{Min $\chi^2$} & \colhead{Age (Myr)} & \colhead{Parallax (mas)} & \colhead{$M_1$ ($M_{\odot}$)} & \colhead{$M_2$ ($M_{\odot}$)} & \colhead{$M_3$ ($M_{\odot}$)}}
\startdata
AaAb-B & Flat & BHAC15 & $2.8$ & $23.6_{-1.9}^{+1.9}$ & $12.72_{-0.13}^{+0.11}$ & $0.906_{-0.007}^{+0.006}$ & $0.881_{-0.008}^{+0.009}$ & $0.171_{-0.046}^{+0.056}$ \\
 & & Padova & $2.8$ & $22.8_{-1.8}^{+1.8}$ & $12.66_{-0.12}^{+0.11}$ & $0.916_{-0.008}^{+0.008}$ & $0.888_{-0.008}^{+0.008}$ & $0.181_{-0.046}^{+0.051}$ \\
 & & Seiss & $5.9$ & $24.5_{-1.8}^{+1.8}$ & $12.41_{-0.13}^{+0.10}$ & $0.984_{-0.010}^{+0.009}$ & $0.955_{-0.009}^{+0.009}$ & $0.169_{-0.047}^{+0.062}$ \\
 & & Yonsei-Yale & $6.3$ & $25.9_{-1.9}^{+2.0}$ & $12.69_{-0.20}^{+0.16}$ & $0.905_{-0.011}^{+0.009}$ & $0.875_{-0.019}^{+0.016}$ & $0.193_{-0.062}^{+0.083}$ \\
 & Gaia DR2 & BHAC15 & $2.7$ & $23.4_{-1.8}^{+1.8}$ & $12.59_{-0.11}^{+0.12}$ & $0.907_{-0.007}^{+0.006}$ & $0.888_{-0.009}^{+0.008}$ & $0.222_{-0.051}^{+0.051}$ \\
 & & Padova & $2.7$ & $22.8_{-1.7}^{+1.7}$ & $12.57_{-0.10}^{+0.11}$ & $0.916_{-0.007}^{+0.008}$ & $0.893_{-0.008}^{+0.008}$ & $0.218_{-0.047}^{+0.047}$ \\
 & & Seiss & $5.5$ & $24.4_{-1.8}^{+1.7}$ & $12.34_{-0.13}^{+0.12}$ & $0.984_{-0.010}^{+0.009}$ & $0.958_{-0.009}^{+0.009}$ & $0.200_{-0.057}^{+0.061}$ \\
 & & Yonsei-Yale & $6.6$ & $25.4_{-1.8}^{+1.9}$ & $12.48_{-0.12}^{+0.15}$ & $0.906_{-0.010}^{+0.008}$ & $0.888_{-0.014}^{+0.011}$ & $0.276_{-0.065}^{+0.056}$ \\
A-B & Flat & BHAC15 & $5.8$ & $23.7_{-1.7}^{+1.8}$ & $13.16_{-0.04}^{+0.04}$ & $0.904_{-0.007}^{+0.006}$ & $0.862_{-0.006}^{+0.006}$ & \nodata \\
 & & Padova & $5.1$ & $22.2_{-2.0}^{+2.0}$ & $13.13_{-0.04}^{+0.04}$ & $0.915_{-0.008}^{+0.008}$ & $0.866_{-0.007}^{+0.007}$ & \nodata \\
 & & Seiss & $5.9$ & $24.1_{-1.7}^{+1.7}$ & $12.79_{-0.04}^{+0.04}$ & $0.984_{-0.009}^{+0.009}$ & $0.939_{-0.007}^{+0.007}$ & \nodata \\
 & & Yonsei-Yale & $6.9$ & $26.7_{-1.9}^{+1.8}$ & $13.24_{-0.07}^{+0.06}$ & $0.897_{-0.012}^{+0.011}$ & $0.840_{-0.011}^{+0.015}$ & \nodata \\
 & Gaia DR2 & BHAC15 & $5.5$ & $23.2_{-1.7}^{+1.7}$ & $13.14_{-0.03}^{+0.04}$ & $0.907_{-0.006}^{+0.006}$ & $0.865_{-0.006}^{+0.006}$ & \nodata \\
 & & Padova & $5.2$ & $21.9_{-1.9}^{+1.8}$ & $13.10_{-0.04}^{+0.04}$ & $0.918_{-0.008}^{+0.008}$ & $0.870_{-0.007}^{+0.007}$ & \nodata \\
 & & Seiss & $5.9$ & $23.8_{-1.6}^{+1.6}$ & $12.78_{-0.04}^{+0.04}$ & $0.987_{-0.009}^{+0.008}$ & $0.941_{-0.007}^{+0.007}$ & \nodata \\
 & & Yonsei-Yale & $7.2$ & $25.4_{-1.8}^{+1.8}$ & $13.17_{-0.06}^{+0.06}$ & $0.907_{-0.011}^{+0.009}$ & $0.853_{-0.014}^{+0.014}$ & \nodata \\
A-BaBb & Flat & BHAC15 & $10.6$ & $22.9_{-1.7}^{+1.7}$ & $12.86_{-0.05}^{+0.04}$ & $0.916_{-0.006}^{+0.007}$ & $0.861_{-0.006}^{+0.006}$ & $0.113_{-0.010}^{+0.020}$ \\
 & & Padova & $11.8$ & $20.3_{-2.1}^{+2.4}$ & $12.84_{-0.05}^{+0.04}$ & $0.929_{-0.007}^{+0.008}$ & $0.865_{-0.008}^{+0.008}$ & $0.108_{-0.006}^{+0.013}$ \\
 & & Seiss & $10.9$ & $23.6_{-1.6}^{+1.6}$ & $12.51_{-0.06}^{+0.05}$ & $0.998_{-0.008}^{+0.008}$ & $0.940_{-0.006}^{+0.006}$ & $0.119_{-0.014}^{+0.028}$ \\
 & & Yonsei-Yale & $10.6$ & $26.0_{-1.7}^{+1.7}$ & $12.84_{-0.11}^{+0.08}$ & $0.920_{-0.010}^{+0.009}$ & $0.847_{-0.011}^{+0.013}$ & $0.132_{-0.024}^{+0.047}$ \\
 & Gaia DR2 & BHAC15 & $12.4$ & $22.7_{-1.7}^{+1.7}$ & $12.83_{-0.07}^{+0.05}$ & $0.918_{-0.006}^{+0.007}$ & $0.862_{-0.006}^{+0.006}$ & $0.121_{-0.015}^{+0.028}$ \\
 & & Padova & $11.3$ & $20.2_{-2.0}^{+2.3}$ & $12.82_{-0.05}^{+0.04}$ & $0.932_{-0.007}^{+0.008}$ & $0.867_{-0.008}^{+0.008}$ & $0.111_{-0.008}^{+0.018}$ \\
 & & Seiss & $9.6$ & $23.4_{-1.7}^{+1.6}$ & $12.49_{-0.07}^{+0.05}$ & $1.000_{-0.008}^{+0.008}$ & $0.941_{-0.006}^{+0.006}$ & $0.125_{-0.019}^{+0.034}$ \\
 & & Yonsei-Yale & $12.0$ & $25.5_{-1.6}^{+1.6}$ & $12.71_{-0.16}^{+0.13}$ & $0.928_{-0.010}^{+0.010}$ & $0.849_{-0.011}^{+0.013}$ & $0.183_{-0.054}^{+0.071}$ \\
\enddata
\end{deluxetable*}

\begin{figure}
\label{tab:fluxwavelength}
\includegraphics[width=\columnwidth]{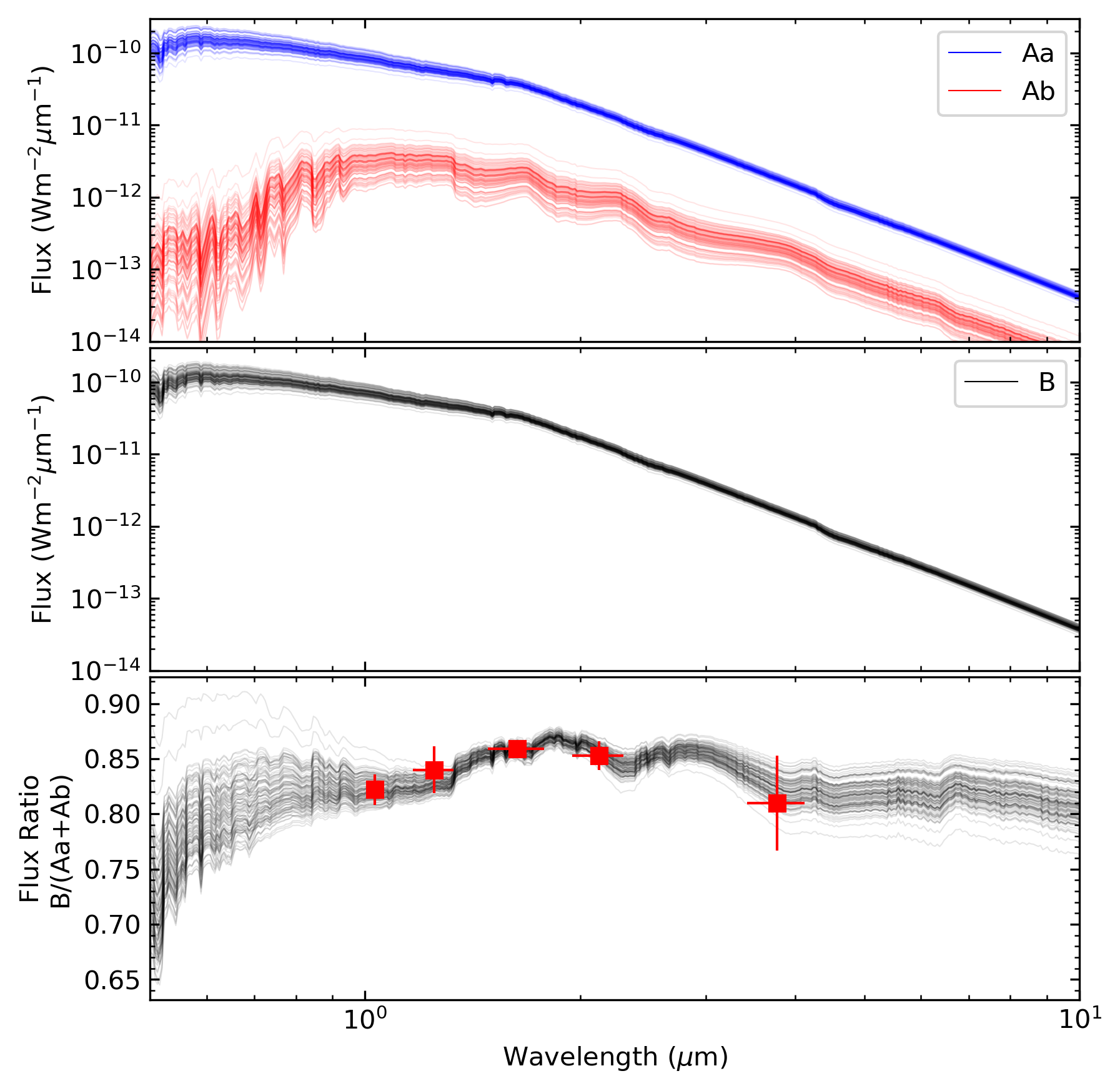}
\caption{Model fluxes of stars Aa and Ab (top panel) and B (middle panel), and the flux ratio of the resolved pair AaAb and B (bottom panel) as a function of wavelength from the Padova evolutionary model for a hundred samples drawn from our MCMC analysis. The red squares represent the flux ratio between the two components measured in the Keck/NIRC2 images taken on 2020 July 24.}
\label{fig:fluxwave}
\end{figure}

The results of our experiment are reported in Table \ref{tab:fluxwavelength}. We find a best fit to the BHAC15 and Padova model grids in the AaAb-B configurations, with a minimum $\chi^2$ of 2.8, excluding the contribution from the visual orbit fit, compared with $\sim$5 for the A-B configuration, and 10--12 for the A-BaBb configuration. The Seiss and Yonsei-Yale models show similar goodness of fit for the A-B and AaAb-B configurations, with the A-BaBb configuration being more strongly disfavoured. The fits with and without a prior on the parallax give a generally similar goodness of fit, due in part to the relatively large uncertainty of the Gaia DR2 parallax measurement.
This joint fit of the astrometric and photometric measurements of the CD-27 system provides some evidence of an unresolved companion to the A component within the CD-27 system, although the improvement in the goodness of fit does depend on which evolutionary model is used. We compute the component masses for the AaAb-B system architecture by taking the weighted average of the four model-dependent results and find $M_{\rm Aa} = 0.923\pm0.008$\,$M_\odot$, $M_{\rm Ab} = 0.177\pm0.055$\,$M_\odot$ and $M_{\rm B} = 0.903\pm0.009$\,$M_\odot$. The age of the system with this method is $24.1 \pm 1.9$\,Myr and the parallax is $12.61 \pm 0.13$\,mas.

\subsection{An Overestimated Distance}
\label{sec:gaia_wrong}
Estimating the system mass and absolute magnitudes of both components requires a reliable measurement of the system parallax. The discrepancy between measured and predicted system mass could be explained by decreasing the distance to the star. This would cause a decrease in the system mass due to a decrease in the semi-major axis, and would decrease the absolute magnitudes of the two components. This scenario was considered in Section~\ref{Flux Ratio Fitting}, where we found no significant difference in the goodness of fit when comparing fits where the parallax was constrained based on the \textit{Gaia} DR2 measurement to those where it was allowed to float freely. Nevertheless, the large uncertainty on the parallax measurement suggests this is a plausible explanation for the observed discrepancy.

\begin{figure}
\includegraphics[width=\columnwidth]{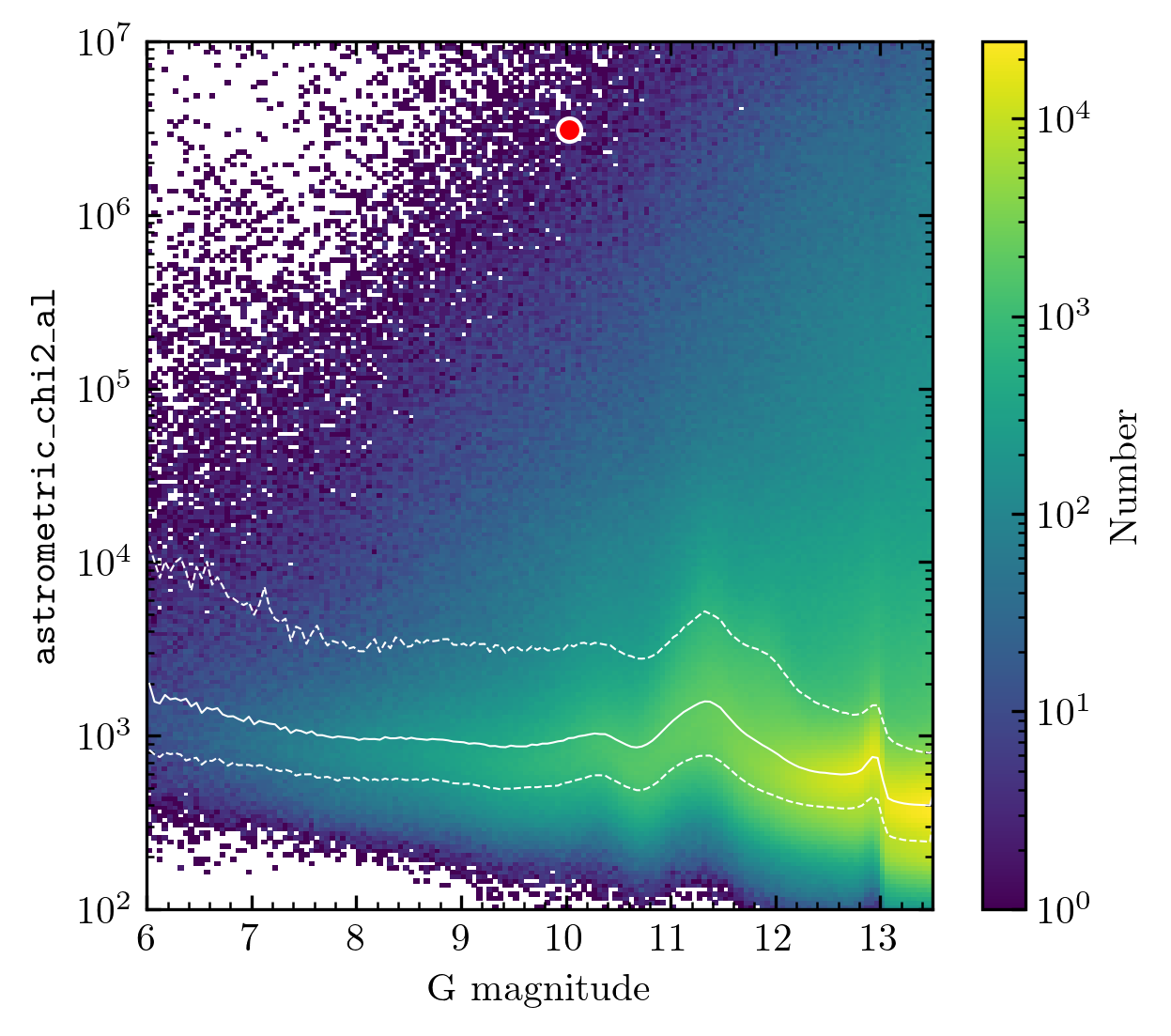}
\caption{The $\chi^2$ reported in the {\it Gaia} DR3 catalogue for the fit to the astrometric measurements of CD-27 11535 (red point) compared to stars of a range of magnitudes. CD-27 has a $\chi^2$ larger than 99.5\,\% of stars within 0.1 mag. This is consistent with the variability of CD-27 biasing the Gaia DR2 5-parameter fit, casting doubt on the accuracy of the parallax.}
\label{fig:gaiadr3chisq}
\end{figure}

The astrometric parameters of the CD-27 system reported in \textit{Gaia} DR2 are derived from a five-parameter  astrometric fit (position, proper motion, parallax) based on 111 measurements taken over the course of almost two years. All of the sources within \textit{Gaia} DR2 were treated as single stars, assuming constant linear motion over the time-span of the mission. Stars that deviate from linear motion for whatever reason will still be fit using the same five-parameter model, but the goodness of fit will suffer due to the incorrect assumption of linear motion. For CD-27, \textit{Gaia} DR2 reports a unit weight error of $u=4.8$, well above the typical value of $u=1.4$ for stars with $G\sim10$ \citep{GAIADATADR2}. This poor goodness of fit is reflected in the unusually large uncertainties on the five fitted astrometric parameters. The goodness of fit is significantly worse in the subsequent \textit{Gaia} Data Release 3 (DR3; \citealp{GAIAEDR3:2021}), as shown in Figure~\ref{fig:gaiadr3chisq}, and only a two-parameter solution was presented. The star was not listed in the non-single star supplement in DR3, suggesting that an astrometric binary solution was attempted, but rejected.

We consider two plausible causes for this apparent deviation from linear motion that could explain the poor goodness of fit in both \textit{Gaia} DR2 and DR3, and that could have biased the parallax measurement reported in \textit{Gaia} DR2. We first consider the effect of orbital motion of the two known components over the course of the \textit{Gaia} observations. We use the orbit fit presented in Section~\ref{Orbit Fitting} and the flux ratios in Table~\ref{tab:astrometry}, under the assumption that $\Delta G \sim \Delta z$, to predict the motion of the photocenter as observed by \textit{Gaia} from 2014 Jul 25 to 2016 May 23. We predict almost linear motion in the declination direction with little change in right ascension. This would manifest itself as a small change to the measured proper motion of the system rather than inducing significant non-linear motion. It is unlikely that the orbital motion of the two resolved components could explain the poor goodness of fit in \textit{Gaia} DR2.

Another source of non-linear astrometric motion in long-period binary systems is the variability of one or both of the components. So-called variability-induced movers (VIMs; \citealp{Wielen:1996}) are systems in which the periodic variability of an unresolved binary causes the photocenter to move in a characteristic fashion. Several examples have been identified in the {\it Hipparcos} and {\it Gaia} DR3 catalogues. Using the measured $\sim$15\% amplitude of the $V$-band variability of this system \citep{Kiraga:2012}, we estimate that the variability can shift the {\it Gaia} photocenter relative to the barycenter by as much as 10\,mas between minima and maxima if the variability is confined to one component. Given this is comparable to the amplitude of the parallax signal, it is likely that significant variability could significantly bias the parallax measurements. Unfortunately, without access to the individual {\it Gaia} measurements, it is not possible to attempt to fit the parallax whilst simultaneously modelling this effect. CD-27 is not flagged as a variability-induced mover in {\it Gaia} DR3, and it is not possible to ascertain whether such a solution was attempted.

\section{Limits on the Presence of Additional Companions}
\label{sec:limits}
\subsection{Lower Limit from Spectroscopy}
Our spectroscopic measurements can be used to place a lower limit on the orbital period of an additional companion within the system. A low-mass companion on a very short orbital period would induce a significant change of the radial velocity of the star it orbits, so long as the orbit is not face-on. We search for the effects of a third component within the ARCES spectra by cross-correlating each with the spectrum of HIP 41277, a K8 star with a comparable $v\sin i$ to CD-27. We use eight orders of the ARCES spectrum covering $\sim$5500\,\AA~to $\sim$6700\,\AA, each with several lines and high signal-to-noise. The resulting cross-correlation function has a single peak for each epoch, and we do not measure a significant variation of the radial velocity.
To explore the effects of relative radial velocity and rotational broadening on the cross-correlation function, we generate a synthetic binary spectrum with varying radial velocity offsets and rotational velocities which we cross-correlate with the spectrum of HIP 41277. We generate the simulated binary by combining two copies of a spectrum of a K5 star we observed as an RV standard (HIP 48331; \citealp{Soubiran:2013}). Each spectrum is rotationally-broadened and scaled based on the predicted optical flux ratio for the two stars. We sum and normalize the blended spectrum, which is then cross-correlated with HIP 41277. This process is repeated for a grid of radial velocity offsets (six between 0\,km\,s$^{-1}$ and 40\,km\,s$^{-1}$) and $v\sin i$ (four between 5\,km\,s$^{-1}$ and 20\,km\,s$^{-1}$.
The resulting cross-correlation functions are shown in Figure \ref{fig:apo_ccf}. At small delta RV ($\lesssim25$\,km\,s$^{-1}$) and $v\sin i$ ($\lesssim5$\,km\,s$^{-1}$), the cross-correlation function has a single peak. When lines are broadened by rotation ($v\sin i \gtrsim 20$\,km\,s$^{-1}$), a larger delta RV ($\gtrsim40$\,km\,s$^{-1}$) is required to resolve them. We identify the delta RV at which we detect two visually distinct peaks in the cross-correlation function, indicating a detectable separation between the primary and secondary absorption lines. Typical $v\sin i$ for K-type stars in the $\beta$~Pic moving group are $10$ to $13$\,km\,s$^{-1}$ \citep{Zuckerman:2001}. For $v\sin i = 10$\,km\,s$^{-1}$, the smallest delta RV at which we see two distinct peaks is $\sim 25$\,km\,s$^{-1}$. For $v\sin i = 13$\,km\,s$^{-1}$, the smallest delta RV at which we see two distinct peaks is $\sim30$\,km\,s$^{-1}$. The predicted delta RV from our orbit fit for A and B alone is $\sim5$\,km\,s$^{-1}$. Thus, there can be an additional delta radial velocity $\gtrsim25$\,km\,s$^{-1}$ caused by the orbit of a companion around either component that would remain undetectable in our ARCES spectra. This upper limit on the radial velocity, assuming zero eccentricity and a 90-degree inclination angle and the masses given in Table \ref{tab:possibleconfigs} row 2, corresponds to a semi-major axis larger than $0.04$\,au (orbital period $> 0.01$\,yr).

\subsection{Upper Limit from Astrometry}
\begin{figure}[h]
\includegraphics[scale=0.3]{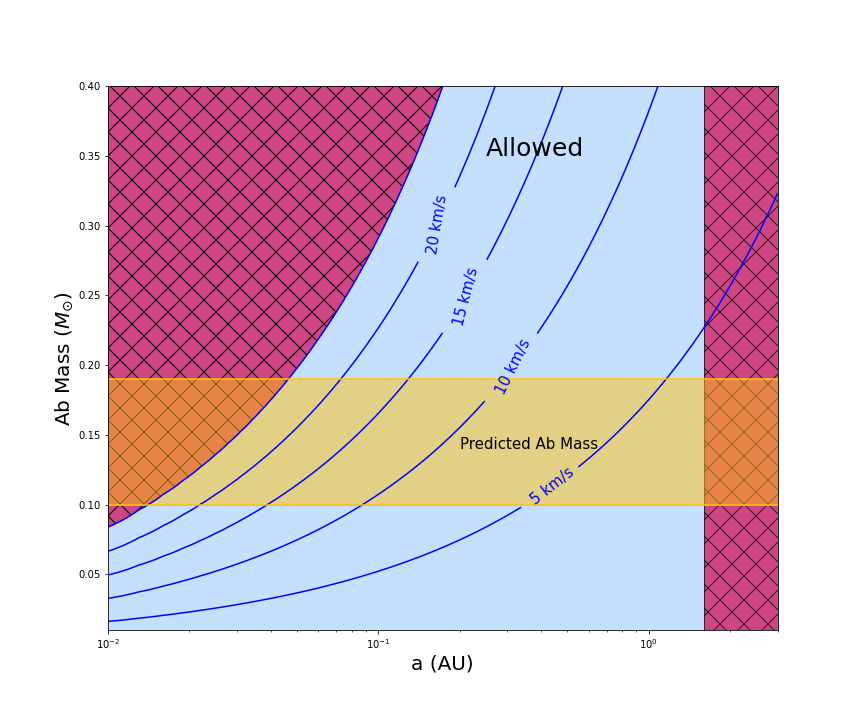}
\caption{The pink, hatched regions indicate mass-separation combinations that are not allowed for Ab. On the left, the semi-major axis is limited by the maximum possible radial velocity ($\sim$25\,km\,s$^{-1}$) of Aa induced by Ab over a range of Ab masses. The range of predicted Ab masses is highlighted in yellow. On the right, the lack of a detectable wobble in the astrometric measurements allows us to draw a conservative upper limit of $\sim$1.6\,au for the subsystem semi-major axis from the largest separation residual ($\sim$2\,mas) and dynamical arguments.}
\label{fig:allowed}
\end{figure}

As there is no obvious Keplerian signal in the residuals of the orbit fit to our astrometric measurements (Fig.~\ref{fig:OrbitSepPa}, second and fourth panels), we can place an upper limit on the semi-major axis of an inner binary. We make the conservative assumption that an orbit with a photocenter semi-amplitude of 2\,mas in the near-infrared would have easily been detected. Using the predicted masses and fluxes of the three components given for the AaAb-B configuration in Table~\ref{tab:possibleconfigs}, we convert this maximum photocenter semi-major axis into an upper limit on the total semi-major axis of the inner subsystem. We compute the reduced mass $B=M_2/\left(M_1+M_2\right)$ and reduced flux $\beta=F_2/\left(F_1+F_2\right)$, from which we can calculate the total semi-major axis as $a=a_p/(B-\beta)$. We find a conservative upper limit for the semi-major axis of the inner binary of $\sim$1.6\,au (orbital period $<2.0$\,yr). The allowed phase space for the semi-major axis of an inner binary is shown in Figure \ref{fig:allowed}.

\section{Conclusion}
\label{sec:conclusion}
We have presented resolved astrometric measurements of the young CD-27 11535 system that we used to refine the visual orbit of the system. Our analysis revealed a significant discrepancy between the total system mass, and the masses of the components estimated for evolutionary models. We explored two scenarios that could explain this discrepancy; either an unresolved companion to either of the two stars, or a biased parallax measurement caused by the photometric variability of either or both of the components. Our joint fit of the visual orbit and resolved photometry provides evidence for a low-mass star present around the A component of the system. Our most favoured configuration for this scenario is a K-type star and M dwarf in a short but unknown orbit (with a semi-major axis between 0.04 and 1.6\,au), with a more distant K-type star in a $\sim$20-year orbit with a semi-major axis of 9.8\,au. This configuration is favoured for the BHAC15 and Padova model grids regardless of whether the parallax is constrained using the {\it Gaia} DR2 measurement or not.

An alternative explanation for the observed discrepancy between the measured and predicted system mass is a bias in the parallax measurement reported in {\it Gaia} DR2. CD-27 has a very poor goodness of fit in both {\it Gaia} DR2 and DR3, so much so that only a two-parameter fit is provided in the DR3. We exclude the possibility that this poor goodness of fit is caused by the photocenter orbit of the known binary; the orbital motion is almost linear over the {\it Gaia} baseline. Instead, it is likely that the photometric variability of one or both of the components is causing a significant motion of the photocenter relative to the barycenter on periods comparable to the 2.9\,d variability measured by \cite{Kiraga:2012}. The amplitude of this variability-induced motion is predicted to be on the same order as the amplitude of the parallax signal. Unfortunately, without the individual {\it Gaia} measurements, it is impossible to re-fit the parallax accounting for this effect.

Currently, the uncertainty regarding the architecture of the system is primarily driven by uncertainty in the parallax. The upcoming {\it Gaia} data releases will contain the individual astrometric measurements made during the five and ten-year duration of the mission, covering a good fraction of the orbital period of the system. With these data in hand, we will be able to jointly fit the systemic motion and parallax of the system, the orbital motion, and motion induced by variability of either or both of the resolved components. With a refined distance estimate in hand, we will be able to re-visit the observed discrepancy between the measured and predicted system mass. In the mean time, interferometric observations from CHARA \citep{2005CHARA} and VLTI \citep{2003VLTI}, as well as high-resolution near-infrared spectroscopic observations, will be able to place improved constraints on the presence of additional companions, complementing the analysis performed here. A complete characterization of the CD-27 system, both in terms of an improved distance determination, and a more detailed search for additional companion, will be necessary before the system can be used as a valuable benchmark for the study of the evolution of young stars.

\begin{acknowledgments}

This work was supported in part by NASA grants NNX14AJ80G, 80NSSC21K0958 (E.L.N. and A.E.P), and 21-ADAP21-0130 (E.L.N. and A.S) and authored by employees of Caltech/IPAC under Contract No. 80GSFC21R0032 with the National Aeronautics and Space Administration. Some of the data presented herein were obtained at the W. M. Keck Observatory, which is operated as a scientific partnership among the California Institute of Technology, the University of California and the National Aeronautics and Space Administration. The Observatory was made possible by the generous financial support of the W. M. Keck Foundation. The authors wish to recognize and acknowledge the very significant cultural role and reverence that the summit of Maunakea has always had within the indigenous Hawaiian community.  We are most fortunate to have the opportunity to conduct observations from this mountain. This work is partially based on observations obtained at the international Gemini Observatory, a program of NSF’s NOIRLab, which is managed by the Association of Universities for Research in Astronomy (AURA) under a cooperative agreement with the National Science Foundation on behalf of the Gemini Observatory partnership: the National Science Foundation (United States), National Research Council (Canada), Agencia Nacional de Investigaci\'{o}n y Desarrollo (Chile), Ministerio de Ciencia, Tecnolog\'{i}a e Innovaci\'{o}n (Argentina), Minist\'{e}rio da Ci\^{e}ncia, Tecnologia, Inova\c{c}\~{o}es e Comunica\c{c}\~{o}es (Brazil), and Korea Astronomy and Space Science Institute (Republic of Korea).
This research has also made use of the VizieR catalogue access tool, CDS, Strasbourg, France (DOI : 10.26093/cds/vizier) and the SIMBAD database, operated at the same location. The original description of the VizieR service was published in 2000, A\&AS 143, 23. Further, this work presents results from the European Space Agency (ESA) space mission Gaia. Gaia data are being processed by the Gaia Data Processing and Analysis Consortium (DPAC). Funding for the DPAC is provided by national institutions, in particular the institutions participating in the Gaia MultiLateral Agreement (MLA). The Gaia mission website is https://www.cosmos.esa.int/gaia. The Gaia archive website is https://archives.esac.esa.int/gaia. Finally, this publication makes use of data products from the Two Micron All Sky Survey, which is a joint project of the University of Massachusetts and the Infrared Processing and Analysis Center/California Institute of Technology, funded by the National Aeronautics and Space Administration and the National Science Foundation.
\end{acknowledgments}
\pagebreak

\bibliography{refs.bib}

\begin{thebibliography}{}
\expandafter\ifx\csname natexlab\endcsname\relax\def\natexlab#1{#1}\fi
\providecommand{\url}[1]{\href{#1}{#1}}
\providecommand{\dodoi}[1]{doi:~\href{http://doi.org/#1}{\nolinkurl{#1}}}
\providecommand{\doeprint}[1]{\href{http://ascl.net/#1}{\nolinkurl{http://ascl.net/#1}}}
\providecommand{\doarXiv}[1]{\href{https://arxiv.org/abs/#1}{\nolinkurl{https://arxiv.org/abs/#1}}}

\bibitem[{{Alonso-Floriano} {et~al.}(2015){Alonso-Floriano}, {Caballero},
  {Cort{\'e}s-Contreras}, {Solano}, \& {Montes}}]{Alonso:2015}
{Alonso-Floriano}, F.~J., {Caballero}, J.~A., {Cort{\'e}s-Contreras}, M.,
  {Solano}, E., \& {Montes}, D. 2015, \aap, 583, A85,
  \dodoi{10.1051/0004-6361/201526795}

\bibitem[{{Baraffe} {et~al.}(2015){Baraffe}, {Homeier}, {Allard}, \&
  {Chabrier}}]{baraffe:2015}
{Baraffe}, I., {Homeier}, D., {Allard}, F., \& {Chabrier}, G. 2015, \aap, 577,
  A42, \dodoi{10.1051/0004-6361/201425481}

\bibitem[{{Bell} {et~al.}(2015){Bell}, {Mamajek}, \& {Naylor}}]{bell:2015}
{Bell}, C.~P.~M., {Mamajek}, E.~E., \& {Naylor}, T. 2015, \mnras, 454, 593,
  \dodoi{10.1093/mnras/stv1981}

\bibitem[{{Blunt} {et~al.}(2017){Blunt}, {Nielsen}, {De Rosa}, {Konopacky},
  {Ryan}, {Wang}, {Pueyo}, {Rameau}, {Marois}, {Marchis}, {Macintosh},
  {Graham}, {Duch{\^e}ne}, \& {Schneider}}]{Blunt:2017}
{Blunt}, S., {Nielsen}, E.~L., {De Rosa}, R.~J., {et~al.} 2017, \aj, 153, 229,
  \dodoi{10.3847/1538-3881/aa6930}

\bibitem[{{Blunt} {et~al.}(2020){Blunt}, {Wang}, {Angelo}, {Ngo}, {Cody}, {De
  Rosa}, {Graham}, {Hirsch}, {Nagpal}, {Nielsen}, {Pearce}, {Rice}, \&
  {Tejada}}]{Blunt2020}
{Blunt}, S., {Wang}, J.~J., {Angelo}, I., {et~al.} 2020, \aj, 159, 89,
  \dodoi{10.3847/1538-3881/ab6663}

\bibitem[{{Bonavita} {et~al.}(2022){Bonavita}, {Gratton}, {Desidera},
  {Squicciarini}, {D'Orazi}, {Zurlo}, {Biller}, {Chauvin}, {Fontanive},
  {Janson}, {Messina}, {Menard}, {Meyer}, {Vigan}, {Avenhaus}, {Asensio
  Torres}, {Beuzit}, {Boccaletti}, {Bonnefoy}, {Brandner}, {Cantalloube},
  {Cheetham}, {Cudel}, {Daemgen}, {Delorme}, {Desgrange}, {Dominik}, {Engler},
  {Feautrier}, {Feldt}, {Galicher}, {Garufi}, {Gasparri}, {Ginski}, {Girard},
  {Grandjean}, {Hagelberg}, {Henning}, {Hunziker}, {Kasper}, {Keppler},
  {Lagadec}, {Lagrange}, {Langlois}, {Lannier}, {Lazzoni}, {Le Coroller},
  {Ligi}, {Lombart}, {Maire}, {Mazevet}, {Mesa}, {Mouillet}, {Moutou},
  {M{\"u}ller}, {Peretti}, {Perrot}, {Petrus}, {Potier}, {Ramos}, {Rickman},
  {Rouan}, {Salter}, {Samland}, {Schmidt}, {Sissa}, {Stolker}, {Szul{\'a}gyi},
  {Turatto}, {Udry}, \& {Wildi}}]{Bonavita:2022}
{Bonavita}, M., {Gratton}, R., {Desidera}, S., {et~al.} 2022, \aap, 663, A144,
  \dodoi{10.1051/0004-6361/202140510}

\bibitem[{{Bressan} {et~al.}(2012){Bressan}, {Marigo}, {Girardi}, {Salasnich},
  {Dal Cero}, {Rubele}, \& {Nanni}}]{bressan:2012}
{Bressan}, A., {Marigo}, P., {Girardi}, L., {et~al.} 2012, \mnras, 427, 127,
  \dodoi{10.1111/j.1365-2966.2012.21948.x}

\bibitem[{Burgasser \& Blake(2009)}]{Burgasser2009}
Burgasser, A.~J., \& Blake, C.~H. 2009, The Astronomical Journal, 137, 4621,
  \dodoi{10.1088/0004-6256/137/6/4621}

\bibitem[{Carlos {et~al.}(2016)Carlos, Nissen, \& Mel{\'{e}}ndez}]{Carlos:2016}
Carlos, M., Nissen, P.~E., \& Mel{\'{e}}ndez, J. 2016, Astronomy \&
  Astrophysics, 587, A100, \dodoi{10.1051/0004-6361/201527478}

\bibitem[{Chauvin {et~al.}(2015)Chauvin, Vigan, Bonnefoy, Desidera, Bonavita,
  Mesa, Boccaletti, Buenzli, Carson, Delorme, Hagelberg, Montagnier, Mordasini,
  Quanz, Segransan, Thalmann, Beuzit, Biller, Covino, Feldt, Girard, Gratton,
  Henning, Kasper, Lagrange, Messina, Meyer, Mouillet, Moutou, Reggiani,
  Schlieder, \& Zurlo}]{Chauvin:2015jy}
Chauvin, G., Vigan, A., Bonnefoy, M., {et~al.} 2015, A{\&}A, 573, A127

\bibitem[{{Correia} {et~al.}(2003){Correia}, {Richichi}, \&
  {Sch{\"o}ller}}]{2003VLTI}
{Correia}, S., {Richichi}, A., \& {Sch{\"o}ller}, M. 2003, \apss, 286, 191,
  \dodoi{10.1023/A:1026124026341}

\bibitem[{{Cutri} {et~al.}(2003){Cutri}, {Skrutskie}, {van Dyk}, {Beichman},
  {Carpenter}, {Chester}, {Cambresy}, {Evans}, {Fowler}, {Gizis}, {Howard},
  {Huchra}, {Jarrett}, {Kopan}, {Kirkpatrick}, {Light}, {Marsh}, {McCallon},
  {Schneider}, {Stiening}, {Sykes}, {Weinberg}, {Wheaton}, {Wheelock}, \&
  {Zacarias}}]{Cutri:2003}
{Cutri}, R.~M., {Skrutskie}, M.~F., {van Dyk}, S., {et~al.} 2003, {2MASS All
  Sky Catalog of point sources.}

\bibitem[{De~Rosa {et~al.}(2020)De~Rosa, Nguyen, Chilcote, Macintosh, Perrin,
  Konopacky, Wang, Duch{\^e}ne, Nielsen, Rameau, Ammons, Bailey, Barman,
  Bulger, Cotten, Doyon, Esposito, Fitzgerald, Follette, Gerard, Goodsell,
  Graham, Greenbaum, Hibon, Hung, Ingraham, Kalas, Larkin, Maire, Marchis,
  Marley, Marois, Metchev, Millar-Blanchaer, Oppenheimer, Palmer, Patience,
  Poyneer, Pueyo, Rajan, Rantakyr{\"o}, Ruffio, Savransky, Schneider,
  Sivaramakrishnan, Song, Soummer, Thomas, Wallace, Ward-Duong, Wiktorowicz, \&
  Wolff}]{DeRosa2020:kd}
De~Rosa, R.~J., Nguyen, M.~M., Chilcote, J., {et~al.} 2020, J. Astron. Telesc.
  Instrum. Syst., 6, 015006

\bibitem[{{Ehrenreich} {et~al.}(2010){Ehrenreich}, {Lagrange}, {Montagnier},
  {Chauvin}, {Galland}, {Beuzit}, \& {Rameau}}]{Ehrenreich2010}
{Ehrenreich}, D., {Lagrange}, A.~M., {Montagnier}, G., {et~al.} 2010, \aap,
  523, A73, \dodoi{10.1051/0004-6361/201014763}

\bibitem[{{Elliott} {et~al.}(2014){Elliott}, {Bayo}, {Melo}, {Torres},
  {Sterzik}, \& {Quast}}]{elliott:2014}
{Elliott}, P., {Bayo}, A., {Melo}, C.~H.~F., {et~al.} 2014, \aap, 568, A26,
  \dodoi{10.1051/0004-6361/201423856}

\bibitem[{{Elliott} {et~al.}(2015){Elliott}, {Hu{\'e}lamo}, {Bouy}, {Bayo},
  {Melo}, {Torres}, {Sterzik}, {Quast}, {Chauvin}, \& {Barrado}}]{Elliot:2015}
{Elliott}, P., {Hu{\'e}lamo}, N., {Bouy}, H., {et~al.} 2015, \aap, 580, A88,
  \dodoi{10.1051/0004-6361/201525794}

\bibitem[{{Foreman-Mackey} {et~al.}(2013){Foreman-Mackey}, {Hogg}, {Lang}, \&
  {Goodman}}]{Foreman-Mackey2013}
{Foreman-Mackey}, D., {Hogg}, D.~W., {Lang}, D., \& {Goodman}, J. 2013, \pasp,
  125, 306, \dodoi{10.1086/670067}

\bibitem[{{Gaia Collaboration} {et~al.}(2018){Gaia Collaboration}, {Brown},
  {Vallenari}, {Prusti}, {de Bruijne}, {Babusiaux}, {Bailer-Jones}, {Biermann},
  {Evans}, {Eyer}, {Jansen}, {Jordi}, {Klioner}, {Lammers}, {Lindegren},
  {Luri}, {Mignard}, {Panem}, {Pourbaix}, {Randich}, {Sartoretti}, {Siddiqui},
  {Soubiran}, {van Leeuwen}, {Walton}, {Arenou}, {Bastian}, {Cropper},
  {Drimmel}, {Katz}, {Lattanzi}, {Bakker}, {Cacciari}, {Casta{\~n}eda},
  {Chaoul}, {Cheek}, {De Angeli}, {Fabricius}, {Guerra}, {Holl}, {Masana},
  {Messineo}, {Mowlavi}, {Nienartowicz}, {Panuzzo}, {Portell}, {Riello},
  {Seabroke}, {Tanga}, {Th{\'e}venin}, {Gracia-Abril}, {Comoretto},
  {Garcia-Reinaldos}, {Teyssier}, {Altmann}, {Andrae}, {Audard},
  {Bellas-Velidis}, {Benson}, {Berthier}, {Blomme}, {Burgess}, {Busso},
  {Carry}, {Cellino}, {Clementini}, {Clotet}, {Creevey}, {Davidson}, {De
  Ridder}, {Delchambre}, {Dell'Oro}, {Ducourant},
  {Fern{\'a}ndez-Hern{\'a}ndez}, {Fouesneau}, {Fr{\'e}mat}, {Galluccio},
  {Garc{\'\i}a-Torres}, {Gonz{\'a}lez-N{\'u}{\~n}ez}, {Gonz{\'a}lez-Vidal},
  {Gosset}, {Guy}, {Halbwachs}, {Hambly}, {Harrison}, {Hern{\'a}ndez},
  {Hestroffer}, {Hodgkin}, {Hutton}, {Jasniewicz}, {Jean-Antoine-Piccolo},
  {Jordan}, {Korn}, {Krone-Martins}, {Lanzafame}, {Lebzelter}, {L{\"o}ffler},
  {Manteiga}, {Marrese}, {Mart{\'\i}n-Fleitas}, {Moitinho}, {Mora}, {Muinonen},
  {Osinde}, {Pancino}, {Pauwels}, {Petit}, {Recio-Blanco}, {Richards},
  {Rimoldini}, {Robin}, {Sarro}, {Siopis}, {Smith}, {Sozzetti}, {S{\"u}veges},
  {Torra}, {van Reeven}, {Abbas}, {Abreu Aramburu}, {Accart}, {Aerts},
  {Altavilla}, {{\'A}lvarez}, {Alvarez}, {Alves}, {Anderson}, {Andrei},
  {Anglada Varela}, {Antiche}, {Antoja}, {Arcay}, {Astraatmadja}, {Bach},
  {Baker}, {Balaguer-N{\'u}{\~n}ez}, {Balm}, {Barache}, {Barata}, {Barbato},
  {Barblan}, {Barklem}, {Barrado}, {Barros}, {Barstow}, {Bartholom{\'e}
  Mu{\~n}oz}, {Bassilana}, {Becciani}, {Bellazzini}, {Berihuete}, {Bertone},
  {Bianchi}, {Bienaym{\'e}}, {Blanco-Cuaresma}, {Boch}, {Boeche}, {Bombrun},
  {Borrachero}, {Bossini}, {Bouquillon}, {Bourda}, {Bragaglia}, {Bramante},
  {Breddels}, {Bressan}, {Brouillet}, {Br{\"u}semeister}, {Brugaletta},
  {Bucciarelli}, {Burlacu}, {Busonero}, {Butkevich}, {Buzzi}, {Caffau},
  {Cancelliere}, {Cannizzaro}, {Cantat-Gaudin}, {Carballo}, {Carlucci},
  {Carrasco}, {Casamiquela}, {Castellani}, {Castro-Ginard}, {Charlot},
  {Chemin}, {Chiavassa}, {Cocozza}, {Costigan}, {Cowell}, {Crifo}, {Crosta},
  {Crowley}, {Cuypers}, {Dafonte}, {Damerdji}, {Dapergolas}, {David}, {David},
  {de Laverny}, {De Luise}, {De March}, {de Martino}, {de Souza}, {de Torres},
  {Debosscher}, {del Pozo}, {Delbo}, {Delgado}, {Delgado}, {Di Matteo},
  {Diakite}, {Diener}, {Distefano}, {Dolding}, {Drazinos}, {Dur{\'a}n},
  {Edvardsson}, {Enke}, {Eriksson}, {Esquej}, {Eynard Bontemps}, {Fabre},
  {Fabrizio}, {Faigler}, {Falc{\~a}o}, {Farr{\`a}s Casas}, {Federici},
  {Fedorets}, {Fernique}, {Figueras}, {Filippi}, {Findeisen}, {Fonti},
  {Fraile}, {Fraser}, {Fr{\'e}zouls}, {Gai}, {Galleti}, {Garabato},
  {Garc{\'\i}a-Sedano}, {Garofalo}, {Garralda}, {Gavel}, {Gavras}, {Gerssen},
  {Geyer}, {Giacobbe}, {Gilmore}, {Girona}, {Giuffrida}, {Glass}, {Gomes},
  {Granvik}, {Gueguen}, {Guerrier}, {Guiraud}, {Guti{\'e}rrez-S{\'a}nchez},
  {Haigron}, {Hatzidimitriou}, {Hauser}, {Haywood}, {Heiter}, {Helmi}, {Heu},
  {Hilger}, {Hobbs}, {Hofmann}, {Holland}, {Huckle}, {Hypki}, {Icardi},
  {Jan{\ss}en}, {Jevardat de Fombelle}, {Jonker}, {Juh{\'a}sz}, {Julbe},
  {Karampelas}, {Kewley}, {Klar}, {Kochoska}, {Kohley}, {Kolenberg},
  {Kontizas}, {Kontizas}, {Koposov}, {Kordopatis}, {Kostrzewa-Rutkowska},
  {Koubsky}, {Lambert}, {Lanza}, {Lasne}, {Lavigne}, {Le Fustec}, {Le
  Poncin-Lafitte}, {Lebreton}, {Leccia}, {Leclerc}, {Lecoeur-Taibi},
  {Lenhardt}, {Leroux}, {Liao}, {Licata}, {Lindstr{\o}m}, {Lister}, {Livanou},
  {Lobel}, {L{\'o}pez}, {Managau}, {Mann}, {Mantelet}, {Marchal}, {Marchant},
  {Marconi}, {Marinoni}, {Marschalk{\'o}}, {Marshall}, {Martino}, {Marton},
  {Mary}, {Massari}, {Matijevi{\v{c}}}, {Mazeh}, {McMillan}, {Messina},
  {Michalik}, {Millar}, {Molina}, {Molinaro}, {Moln{\'a}r}, {Montegriffo},
  {Mor}, {Morbidelli}, {Morel}, {Morris}, {Mulone}, {Muraveva}, {Musella},
  {Nelemans}, {Nicastro}, {Noval}, {O'Mullane}, {Ord{\'e}novic},
  {Ord{\'o}{\~n}ez-Blanco}, {Osborne}, {Pagani}, {Pagano}, {Pailler},
  {Palacin}, {Palaversa}, {Panahi}, {Pawlak}, {Piersimoni}, {Pineau}, {Plachy},
  {Plum}, {Poggio}, {Poujoulet}, {Pr{\v{s}}a}, {Pulone}, {Racero}, {Ragaini},
  {Rambaux}, {Ramos-Lerate}, {Regibo}, {Reyl{\'e}}, {Riclet}, {Ripepi}, {Riva},
  {Rivard}, {Rixon}, {Roegiers}, {Roelens}, {Romero-G{\'o}mez}, {Rowell},
  {Royer}, {Ruiz-Dern}, {Sadowski}, {Sagrist{\`a} Sell{\'e}s}, {Sahlmann},
  {Salgado}, {Salguero}, {Sanna}, {Santana-Ros}, {Sarasso}, {Savietto},
  {Schultheis}, {Sciacca}, {Segol}, {Segovia}, {S{\'e}gransan}, {Shih},
  {Siltala}, {Silva}, {Smart}, {Smith}, {Solano}, {Solitro}, {Sordo}, {Soria
  Nieto}, {Souchay}, {Spagna}, {Spoto}, {Stampa}, {Steele},
  {Steidelm{\"u}ller}, {Stephenson}, {Stoev}, {Suess}, {Surdej}, {Szabados},
  {Szegedi-Elek}, {Tapiador}, {Taris}, {Tauran}, {Taylor}, {Teixeira},
  {Terrett}, {Teyssandier}, {Thuillot}, {Titarenko}, {Torra Clotet}, {Turon},
  {Ulla}, {Utrilla}, {Uzzi}, {Vaillant}, {Valentini}, {Valette}, {van Elteren},
  {Van Hemelryck}, {van Leeuwen}, {Vaschetto}, {Vecchiato}, {Veljanoski},
  {Viala}, {Vicente}, {Vogt}, {von Essen}, {Voss}, {Votruba}, {Voutsinas},
  {Walmsley}, {Weiler}, {Wertz}, {Wevers}, {Wyrzykowski}, {Yoldas},
  {{\v{Z}}erjal}, {Ziaeepour}, {Zorec}, {Zschocke}, {Zucker}, {Zurbach}, \&
  {Zwitter}}]{GAIADATADR2}
{Gaia Collaboration}, {Brown}, A.~G.~A., {Vallenari}, A., {et~al.} 2018, \aap,
  616, A1, \dodoi{10.1051/0004-6361/201833051}

\bibitem[{{Gaia Collaboration} {et~al.}(2021){Gaia Collaboration}, {Brown},
  {Vallenari}, {Prusti}, {de Bruijne}, {Babusiaux}, {Biermann}, {Creevey},
  {Evans}, {Eyer}, {Hutton}, {Jansen}, {Jordi}, {Klioner}, {Lammers},
  {Lindegren}, {Luri}, {Mignard}, {Panem}, {Pourbaix}, {Randich}, {Sartoretti},
  {Soubiran}, {Walton}, {Arenou}, {Bailer-Jones}, {Bastian}, {Cropper},
  {Drimmel}, {Katz}, {Lattanzi}, {van Leeuwen}, {Bakker}, {Cacciari},
  {Casta{\~n}eda}, {De Angeli}, {Ducourant}, {Fabricius}, {Fouesneau},
  {Fr{\'e}mat}, {Guerra}, {Guerrier}, {Guiraud}, {Jean-Antoine Piccolo},
  {Masana}, {Messineo}, {Mowlavi}, {Nicolas}, {Nienartowicz}, {Pailler},
  {Panuzzo}, {Riclet}, {Roux}, {Seabroke}, {Sordo}, {Tanga}, {Th{\'e}venin},
  {Gracia-Abril}, {Portell}, {Teyssier}, {Altmann}, {Andrae}, {Bellas-Velidis},
  {Benson}, {Berthier}, {Blomme}, {Brugaletta}, {Burgess}, {Busso}, {Carry},
  {Cellino}, {Cheek}, {Clementini}, {Damerdji}, {Davidson}, {Delchambre},
  {Dell'Oro}, {Fern{\'a}ndez-Hern{\'a}ndez}, {Galluccio}, {Garc{\'\i}a-Lario},
  {Garcia-Reinaldos}, {Gonz{\'a}lez-N{\'u}{\~n}ez}, {Gosset}, {Haigron},
  {Halbwachs}, {Hambly}, {Harrison}, {Hatzidimitriou}, {Heiter},
  {Hern{\'a}ndez}, {Hestroffer}, {Hodgkin}, {Holl}, {Jan{\ss}en}, {Jevardat de
  Fombelle}, {Jordan}, {Krone-Martins}, {Lanzafame}, {L{\"o}ffler}, {Lorca},
  {Manteiga}, {Marchal}, {Marrese}, {Moitinho}, {Mora}, {Muinonen}, {Osborne},
  {Pancino}, {Pauwels}, {Petit}, {Recio-Blanco}, {Richards}, {Riello},
  {Rimoldini}, {Robin}, {Roegiers}, {Rybizki}, {Sarro}, {Siopis}, {Smith},
  {Sozzetti}, {Ulla}, {Utrilla}, {van Leeuwen}, {van Reeven}, {Abbas}, {Abreu
  Aramburu}, {Accart}, {Aerts}, {Aguado}, {Ajaj}, {Altavilla}, {{\'A}lvarez},
  {{\'A}lvarez Cid-Fuentes}, {Alves}, {Anderson}, {Anglada Varela}, {Antoja},
  {Audard}, {Baines}, {Baker}, {Balaguer-N{\'u}{\~n}ez}, {Balbinot}, {Balog},
  {Barache}, {Barbato}, {Barros}, {Barstow}, {Bartolom{\'e}}, {Bassilana},
  {Bauchet}, {Baudesson-Stella}, {Becciani}, {Bellazzini}, {Bernet}, {Bertone},
  {Bianchi}, {Blanco-Cuaresma}, {Boch}, {Bombrun}, {Bossini}, {Bouquillon},
  {Bragaglia}, {Bramante}, {Breedt}, {Bressan}, {Brouillet}, {Bucciarelli},
  {Burlacu}, {Busonero}, {Butkevich}, {Buzzi}, {Caffau}, {Cancelliere},
  {C{\'a}novas}, {Cantat-Gaudin}, {Carballo}, {Carlucci}, {Carnerero},
  {Carrasco}, {Casamiquela}, {Castellani}, {Castro-Ginard}, {Castro Sampol},
  {Chaoul}, {Charlot}, {Chemin}, {Chiavassa}, {Cioni}, {Comoretto}, {Cooper},
  {Cornez}, {Cowell}, {Crifo}, {Crosta}, {Crowley}, {Dafonte}, {Dapergolas},
  {David}, {David}, {de Laverny}, {De Luise}, {De March}, {De Ridder}, {de
  Souza}, {de Teodoro}, {de Torres}, {del Peloso}, {del Pozo}, {Delbo},
  {Delgado}, {Delgado}, {Delisle}, {Di Matteo}, {Diakite}, {Diener},
  {Distefano}, {Dolding}, {Eappachen}, {Edvardsson}, {Enke}, {Esquej}, {Fabre},
  {Fabrizio}, {Faigler}, {Fedorets}, {Fernique}, {Fienga}, {Figueras},
  {Fouron}, {Fragkoudi}, {Fraile}, {Franke}, {Gai}, {Garabato},
  {Garcia-Gutierrez}, {Garc{\'\i}a-Torres}, {Garofalo}, {Gavras}, {Gerlach},
  {Geyer}, {Giacobbe}, {Gilmore}, {Girona}, {Giuffrida}, {Gomel}, {Gomez},
  {Gonzalez-Santamaria}, {Gonz{\'a}lez-Vidal}, {Granvik},
  {Guti{\'e}rrez-S{\'a}nchez}, {Guy}, {Hauser}, {Haywood}, {Helmi}, {Hidalgo},
  {Hilger}, {H{\l}adczuk}, {Hobbs}, {Holland}, {Huckle}, {Jasniewicz},
  {Jonker}, {Juaristi Campillo}, {Julbe}, {Karbevska}, {Kervella}, {Khanna},
  {Kochoska}, {Kontizas}, {Kordopatis}, {Korn}, {Kostrzewa-Rutkowska},
  {Kruszy{\'n}ska}, {Lambert}, {Lanza}, {Lasne}, {Le Campion}, {Le Fustec},
  {Lebreton}, {Lebzelter}, {Leccia}, {Leclerc}, {Lecoeur-Taibi}, {Liao},
  {Licata}, {Lindstr{\o}m}, {Lister}, {Livanou}, {Lobel}, {Madrero Pardo},
  {Managau}, {Mann}, {Marchant}, {Marconi}, {Marcos Santos}, {Marinoni},
  {Marocco}, {Marshall}, {Martin Polo}, {Mart{\'\i}n-Fleitas}, {Masip},
  {Massari}, {Mastrobuono-Battisti}, {Mazeh}, {McMillan}, {Messina},
  {Michalik}, {Millar}, {Mints}, {Molina}, {Molinaro}, {Moln{\'a}r},
  {Montegriffo}, {Mor}, {Morbidelli}, {Morel}, {Morris}, {Mulone}, {Munoz},
  {Muraveva}, {Murphy}, {Musella}, {Noval}, {Ord{\'e}novic}, {Orr{\`u}},
  {Osinde}, {Pagani}, {Pagano}, {Palaversa}, {Palicio}, {Panahi}, {Pawlak},
  {Pe{\~n}alosa Esteller}, {Penttil{\"a}}, {Piersimoni}, {Pineau}, {Plachy},
  {Plum}, {Poggio}, {Poretti}, {Poujoulet}, {Pr{\v{s}}a}, {Pulone}, {Racero},
  {Ragaini}, {Rainer}, {Raiteri}, {Rambaux}, {Ramos}, {Ramos-Lerate}, {Re
  Fiorentin}, {Regibo}, {Reyl{\'e}}, {Ripepi}, {Riva}, {Rixon}, {Robichon},
  {Robin}, {Roelens}, {Rohrbasser}, {Romero-G{\'o}mez}, {Rowell}, {Royer},
  {Rybicki}, {Sadowski}, {Sagrist{\`a} Sell{\'e}s}, {Sahlmann}, {Salgado},
  {Salguero}, {Samaras}, {Sanchez Gimenez}, {Sanna}, {Santove{\~n}a},
  {Sarasso}, {Schultheis}, {Sciacca}, {Segol}, {Segovia}, {S{\'e}gransan},
  {Semeux}, {Shahaf}, {Siddiqui}, {Siebert}, {Siltala}, {Slezak}, {Smart},
  {Solano}, {Solitro}, {Souami}, {Souchay}, {Spagna}, {Spoto}, {Steele},
  {Steidelm{\"u}ller}, {Stephenson}, {S{\"u}veges}, {Szabados}, {Szegedi-Elek},
  {Taris}, {Tauran}, {Taylor}, {Teixeira}, {Thuillot}, {Tonello}, {Torra},
  {Torra}, {Turon}, {Unger}, {Vaillant}, {van Dillen}, {Vanel}, {Vecchiato},
  {Viala}, {Vicente}, {Voutsinas}, {Weiler}, {Wevers}, {Wyrzykowski}, {Yoldas},
  {Yvard}, {Zhao}, {Zorec}, {Zucker}, {Zurbach}, \& {Zwitter}}]{GAIAEDR3:2021}
---. 2021, \aap, 649, A1, \dodoi{10.1051/0004-6361/202039657}

\bibitem[{Greenbaum {et~al.}(2019)Greenbaum, Cheetham, Sivaramakrishnan,
  Rantakyr{\"o}, Duch{\^e}ne, Tuthill, De~Rosa, Oppenheimer, Macintosh, Ammons,
  Bailey, Barman, Bulger, Cardwell, Chilcote, Cotten, Doyon, Fitzgerald,
  Follette, Gerard, Goodsell, Graham, Hibon, Hung, Ingraham, Kalas, Konopacky,
  Larkin, Maire, Marchis, Marley, Marois, Metchev, Millar-Blanchaer, Morzinski,
  Nielsen, Palmer, Patience, Perrin, Poyneer, Pueyo, Rajan, Rameau, Sadakuni,
  Savransky, Schneider, Song, Soummer, Thomas, Wallace, Wang, Ward-Duong,
  Wiktorowicz, \& Wolff}]{Greenbaum:2019hr}
Greenbaum, A.~Z., Cheetham, A., Sivaramakrishnan, A., {et~al.} 2019, AJ, 157,
  249

\bibitem[{{H{\o}g} {et~al.}(2000){H{\o}g}, {Fabricius}, {Makarov}, {Urban},
  {Corbin}, {Wycoff}, {Bastian}, {Schwekendiek}, \& {Wicenec}}]{Hog:2000}
{H{\o}g}, E., {Fabricius}, C., {Makarov}, V.~V., {et~al.} 2000, \aap, 355, L27

\bibitem[{{Kiraga}(2012)}]{Kiraga:2012}
{Kiraga}, M. 2012, \actaa, 62, 67.
\newblock \doarXiv{1204.3825}

\bibitem[{Lenzen {et~al.}(2003)Lenzen, Hartung, Brandner, Finger, Hubin,
  Lacombe, Lagrange, Lehnert, Moorwood, \& Mouillet}]{Lenzen:2003iu}
Lenzen, R., Hartung, M., Brandner, W., {et~al.} 2003, Proc. SPIE, 4841, 944

\bibitem[{Macintosh {et~al.}(2014)Macintosh, Graham, Ingraham, Konopacky,
  Marois, Perrin, Poyneer, Bauman, Barman, Burrows, Cardwell, Chilcote,
  De~Rosa, Dillon, Doyon, Dunn, Erikson, Fitzgerald, Gavel, Goodsell, Hartung,
  Hibon, Kalas, Larkin, Maire, Marchis, Marley, McBride, Millar-Blanchaer,
  Morzinski, Norton, Oppenheimer, Palmer, Patience, Pueyo, Rantakyro, Sadakuni,
  Saddlemyer, Savransky, Serio, Soummer, Sivaramakrishnan, Song, Thomas,
  Wallace, Wiktorowicz, \& Wolff}]{Macintosh:2014js}
Macintosh, B., Graham, J.~R., Ingraham, P., {et~al.} 2014, PNAS, 111, 12661

\bibitem[{Maire {et~al.}(2014)Maire, Ingraham, De~Rosa, Perrin, Rajan,
  Savransky, Wang, Ruffio, Wolff, Chilcote, Doyon, Graham, Greenbaum,
  Konopacky, Larkin, Macintosh, Marois, Millar-Blanchaer, Patience, Pueyo,
  Sivaramakrishnan, Thomas, \& Weiss}]{Maire:2014gs}
Maire, J., Ingraham, P.~J., De~Rosa, R.~J., {et~al.} 2014, Proc. SPIE, 9147, 85

\bibitem[{{Montet} {et~al.}(2015){Montet}, {Bowler}, {Shkolnik}, {Deck},
  {Wang}, {Horch}, {Liu}, {Hillenbrand}, {Kraus}, \&
  {Charbonneau}}]{montet:2015}
{Montet}, B.~T., {Bowler}, B.~P., {Shkolnik}, E.~L., {et~al.} 2015, \apjl, 813,
  L11, \dodoi{10.1088/2041-8205/813/1/L11}

\bibitem[{{Nielsen} {et~al.}(2016){Nielsen}, {De Rosa}, {Wang}, {Rameau},
  {Song}, {Graham}, {Macintosh}, {Ammons}, {Bailey}, {Barman}, {Bulger},
  {Chilcote}, {Cotten}, {Doyon}, {Duch{\^e}ne}, {Fitzgerald}, {Follette},
  {Greenbaum}, {Hibon}, {Hung}, {Ingraham}, {Kalas}, {Konopacky}, {Larkin},
  {Maire}, {Marchis}, {Marley}, {Marois}, {Metchev}, {Millar-Blanchaer},
  {Oppenheimer}, {Palmer}, {Patience}, {Perrin}, {Poyneer}, {Pueyo}, {Rajan},
  {Rantakyr{\"o}}, {Savransky}, {Schneider}, {Sivaramakrishnan}, {Soummer},
  {Thomas}, {Wallace}, {Ward-Duong}, {Wiktorowicz}, \& {Wolff}}]{Nielsen2016}
{Nielsen}, E.~L., {De Rosa}, R.~J., {Wang}, J., {et~al.} 2016, \aj, 152, 175,
  \dodoi{10.3847/0004-6256/152/6/175}

\bibitem[{{Perrin} {et~al.}(2014){Perrin}, {Maire}, {Ingraham}, {Savransky},
  {Millar-Blanchaer}, {Wolff}, {Ruffio}, {Wang}, {Draper}, {Sadakuni},
  {Marois}, {Rajan}, {Fitzgerald}, {Macintosh}, {Graham}, {Doyon}, {Larkin},
  {Chilcote}, {Goodsell}, {Palmer}, {Labrie}, {Beaulieu}, {De Rosa},
  {Greenbaum}, {Hartung}, {Hibon}, {Konopacky}, {Lafreniere}, {Lavigne},
  {Marchis}, {Patience}, {Pueyo}, {Rantakyr{\"o}}, {Soummer},
  {Sivaramakrishnan}, {Thomas}, {Ward-Duong}, \& {Wiktorowicz}}]{Perrin:2014}
{Perrin}, M.~D., {Maire}, J., {Ingraham}, P., {et~al.} 2014, in \procspie, Vol.
  9147, Ground-based and Airborne Instrumentation for Astronomy V, 91473J,
  \dodoi{10.1117/12.2055246}

\bibitem[{Rousset {et~al.}(2003)Rousset, Lacombe, Puget, Hubin, Gendron, Fusco,
  Arsenault, Charton, Feautrier, Gigan, Kern, Lagrange, Madec, Mouillet,
  Rabaud, Rabou, Stadler, \& Zins}]{Rousset:2003hh}
Rousset, G., Lacombe, F., Puget, P., {et~al.} 2003, Proc. SPIE, 4839, 140

\bibitem[{{Schlieder} {et~al.}(2016){Schlieder}, {Skemer}, {Maire}, {Desidera},
  {Hinz}, {Skrutskie}, {Leisenring}, {Bailey}, {Defr{\`e}re}, {Esposito},
  {Strassmeier}, {Weber}, {Biller}, {Bonnefoy}, {Buenzli}, {Close}, {Crepp},
  {Eisner}, {Hofmann}, {Henning}, {Morzinski}, {Schertl}, {Weigelt}, \&
  {Woodward}}]{schlieder:2016}
{Schlieder}, J.~E., {Skemer}, A.~J., {Maire}, A.-L., {et~al.} 2016, \apj, 818,
  1, \dodoi{10.3847/0004-637X/818/1/1}

\bibitem[{{Service} {et~al.}(2016){Service}, {Lu}, {Campbell}, {Sitarski},
  {Ghez}, \& {Anderson}}]{Service2016}
{Service}, M., {Lu}, J.~R., {Campbell}, R., {et~al.} 2016, \pasp, 128, 095004,
  \dodoi{10.1088/1538-3873/128/967/095004}

\bibitem[{{Shkolnik} {et~al.}(2017){Shkolnik}, {Allers}, {Kraus}, {Liu}, \&
  {Flagg}}]{Shkolnik:2017}
{Shkolnik}, E.~L., {Allers}, K.~N., {Kraus}, A.~L., {Liu}, M.~C., \& {Flagg},
  L. 2017, \aj, 154, 69, \dodoi{10.3847/1538-3881/aa77fa}

\bibitem[{{Siess} {et~al.}(2000){Siess}, {Dufour}, \& {Forestini}}]{siess:2000}
{Siess}, L., {Dufour}, E., \& {Forestini}, M. 2000, \aap, 358, 593

\bibitem[{{Song} {et~al.}(2012){Song}, {Zuckerman}, \& {Bessell}}]{Song:2012}
{Song}, I., {Zuckerman}, B., \& {Bessell}, M.~S. 2012, \aj, 144, 8,
  \dodoi{10.1088/0004-6256/144/1/8}

\bibitem[{{Soubiran} {et~al.}(2013){Soubiran}, {Jasniewicz}, {Chemin}, {Crifo},
  {Udry}, {Hestroffer}, \& {Katz}}]{Soubiran:2013}
{Soubiran}, C., {Jasniewicz}, G., {Chemin}, L., {et~al.} 2013, \aap, 552, A64,
  \dodoi{10.1051/0004-6361/201220927}

\bibitem[{{Spada} {et~al.}(2013){Spada}, {Demarque}, {Kim}, \&
  {Sills}}]{spada:2013}
{Spada}, F., {Demarque}, P., {Kim}, Y.-C., \& {Sills}, A. 2013, \apj, 776, 87,
  \dodoi{10.1088/0004-637X/776/2/87}

\bibitem[{{Stanford-Moore} {et~al.}(2020){Stanford-Moore}, {Nielsen}, {De
  Rosa}, {Macintosh}, \& {Czekala}}]{Stanford-Moore:2020}
{Stanford-Moore}, S.~A., {Nielsen}, E.~L., {De Rosa}, R.~J., {Macintosh}, B.,
  \& {Czekala}, I. 2020, \apj, 898, 27, \dodoi{10.3847/1538-4357/ab9a35}

\bibitem[{{ten Brummelaar} {et~al.}(2005){ten Brummelaar}, {McAlister},
  {Ridgway}, {Bagnuolo}, {Turner}, {Sturmann}, {Sturmann}, {Berger}, {Ogden},
  {Cadman}, {Hartkopf}, {Hopper}, \& {Shure}}]{2005CHARA}
{ten Brummelaar}, T.~A., {McAlister}, H.~A., {Ridgway}, S.~T., {et~al.} 2005,
  \apj, 628, 453, \dodoi{10.1086/43072910.48550/arXiv.astro-ph/0504082}

\bibitem[{Tokovinin(2023)}]{Priv_Comm_tokovinin}
Tokovinin, A. 2023, {Private Communication}

\bibitem[{Tokovinin {et~al.}(2018)Tokovinin, Mason, Hartkopf, Mendez, \&
  Horch}]{Tokovinin:2018}
Tokovinin, A., Mason, B.~D., Hartkopf, W.~I., Mendez, R.~A., \& Horch, E.~P.
  2018, The Astronomical Journal, 155, 235, \dodoi{10.3847/1538-3881/aabf8d}

\bibitem[{Tokovinin {et~al.}(2022)Tokovinin, Mason, Mendez, \&
  Costa}]{Tokovinin:2022}
Tokovinin, A., Mason, B.~D., Mendez, R.~A., \& Costa, E. 2022, The Astronomical
  Journal, 164, 58, \dodoi{10.3847/1538-3881/ac78e7}

\bibitem[{Tokovinin {et~al.}(2020)Tokovinin, Mason, Mendez, Costa, \&
  Horch}]{Tokovinin:2020}
Tokovinin, A., Mason, B.~D., Mendez, R.~A., Costa, E., \& Horch, E.~P. 2020,
  The Astronomical Journal, 160, 7, \dodoi{10.3847/1538-3881/ab91c1}

\bibitem[{Tokovinin {et~al.}(2021)Tokovinin, Mason, Mendez, Costa, Mann, \&
  Henry}]{Tokovinin:2021}
Tokovinin, A., Mason, B.~D., Mendez, R.~A., {et~al.} 2021, The Astronomical
  Journal, 162, 41, \dodoi{10.3847/1538-3881/ac00bd}

\bibitem[{Tokovinin {et~al.}(2019)Tokovinin, Mason, Mendez, Horch, \&
  Brice{\~{n} }o}]{Tokovinin:2019}
Tokovinin, A., Mason, B.~D., Mendez, R.~A., Horch, E.~P., \& Brice{\~{n} }o, C.
  2019, The Astronomical Journal, 158, 48, \dodoi{10.3847/1538-3881/ab24e4}

\bibitem[{{Torres} {et~al.}(2006){Torres}, {Quast}, {da Silva}, {de La Reza},
  {Melo}, \& {Sterzik}}]{Torres:2006}
{Torres}, C.~A.~O., {Quast}, G.~R., {da Silva}, L., {et~al.} 2006, \aap, 460,
  695, \dodoi{10.1051/0004-6361:20065602}

\bibitem[{{Wang} {et~al.}(2003){Wang}, {Hildebrand}, {Hobbs}, {Heimsath},
  {Kelderhouse}, {Loewenstein}, {Lucero}, {Rockosi}, {Sandford}, {Sundwall},
  {Thorburn}, \& {York}}]{Wang:2003}
{Wang}, S.-i., {Hildebrand}, R.~H., {Hobbs}, L.~M., {et~al.} 2003, in Society
  of Photo-Optical Instrumentation Engineers (SPIE) Conference Series, Vol.
  4841, Instrument Design and Performance for Optical/Infrared Ground-based
  Telescopes, ed. M.~{Iye} \& A.~F.~M. {Moorwood}, 1145--1156,
  \dodoi{10.1117/12.461447}

\bibitem[{{Weise} {et~al.}(2010){Weise}, {Launhardt}, {Setiawan}, \&
  {Henning}}]{Weise:2010}
{Weise}, P., {Launhardt}, R., {Setiawan}, J., \& {Henning}, T. 2010, \aap, 517,
  A88, \dodoi{10.1051/0004-6361/201014453}

\bibitem[{{Wielen}(1996)}]{Wielen:1996}
{Wielen}, R. 1996, \aap, 314, 679

\bibitem[{{Yelda} {et~al.}(2011){Yelda}, {Lu}, {Ghez}, {Clarkson}, {Anderson},
  {Do}, \& {Matthews}}]{Yelda2011}
{Yelda}, S., {Lu}, J.~R., {Ghez}, A.~M., {et~al.} 2011, \apj, 726, 114,
  \dodoi{10.1088/0004-637X/726/2/114}

\bibitem[{{Zuckerman} \& {Song}(2004)}]{Zuckerman:2004}
{Zuckerman}, B., \& {Song}, I. 2004, \araa, 42, 685,
  \dodoi{10.1146/annurev.astro.42.053102.134111}

\bibitem[{{Zuckerman} {et~al.}(2001){Zuckerman}, {Song}, {Bessell}, \&
  {Webb}}]{Zuckerman:2001}
{Zuckerman}, B., {Song}, I., {Bessell}, M.~S., \& {Webb}, R.~A. 2001, \apjl,
  562, L87, \dodoi{10.1086/337968}

\end{thebibliography}
\begin{appendix}
Table \ref{tab:observing_log} presents the observing log of all epochs for CD-27 and calibrators for which we reduced data. Table \ref{fig:orbitcorner} contains all astrometric and photometric measurements of the star - both our own and those found in the literature - as well as the residuals compared to the median orbit. Figure \ref{fig:orbitcorner} shows the corner plot of the orbital parameters from the initial \texttt{orbitize!} fit and reveals convergence. Figure \ref{fig:cornerpadova} displays the covariance between the system age, parallax and component masses from the joint fit in section \ref{sec:scenarios} using the evolutionary model and system architecture with the lowest $\chi^2$ - AaAb-B and Padova PAR-SEC. Figure \ref{fig:apo_ccf} shows the cross-correlation function between one observed spectrum of CD-27 and HIP 41277, as well as a simulated binary and HIP 41277 for various $v\sin i$ and $\Delta$ RV.

\begin{deluxetable*}{cccccccccc}[b]
\tabletypesize{\scriptsize}
\tablecaption{\label{tab:observing_log}Observing log of CD-27 11535 and suitable PSF calibrators}
\tablewidth{0pt}
\tablehead{
\colhead{UT Start} & \colhead{Target} & \colhead{Camera} & \colhead{Filter} & \colhead{$t_{\rm dit}$} & \colhead{$n_{\rm dit}$} & \colhead{$n_{\rm exp}$} & \colhead{DIMM seeing} & \colhead{Airmass} & \colhead{Program ID}\\
&&&&(s)&&&(asec)}
\startdata
\multicolumn{10}{c}{VLT/NaCo}\\
\hline
\hline
2006-05-27T04:40:28 & CD-27 11535           & S27 & IB2.27 & 1.2 & 50  & 12 & 0.99--1.34 & 1.01--1.03 & 077.C-0483\\
2006-05-27T01:23:21 & 1RXS J125608.8-692652 & S27 & IB2.27 & 50  & 1.2 & 16 & 1.19--1.62 & 1.41       & 077.C-0483\\
2006-05-27T02:00:28 & HD 112245             & S27 & IB2.27 & 75  & 0.8 & 13 & 1.08--1.95 & 1.26--1.27 & 077.C-0483\\
2006-05-27T06:57:01 & CD-54 7336            & S27 & IB2.27 & 75  & 0.8 & 12 & 0.93--1.32 & 1.18--1.20 & 077.C-0483\\
\hline
2008-05-18T04:17:54 & CD-27 11535     & S27 & IB2.27 & 60  & 1.0 & 12 & 0.67--1.23 & 1.08--1.11 & 081.C-0825\\
2008-05-19T06:22:59 & GSC 06242-00004 & S27 & IB2.27 & 150 & 0.4 & 12 & 0.70--0.92 & 1.00--1.01 & 081.C-0825\\
2008-05-19T05:57:49 & TYC 6234-1287-1 & S27 & IB2.27 & 40  & 1.5 & 12 & 0.59--0.72 & 1.01       & 081.C-0825\\
2008-05-17T03:27:16 & CD-25 11504     & S27 & IB2.27 & 100 & 0.6 & 12 & 0.65--0.81 & 1.09--1.12 & 081.C-0825\\
2008-05-17T03:50:14 & V* V2505 Oph    & S27 & IB2.27 & 120 & 0.5 & 12 & 0.59--0.67 & 1.07--1.09 & 081.C-0825\\
\hline
2009-06-01T06:13:30 & CD-27 11535             & S13 & $K_{\rm s}$/ND & 5 & 2 & 9 & 1.32--1.84 & 1.03       & 083.C-0659\\
2009-06-01T00:27:49 & 2MASS J12205449-6457242 & S13 & $K_{\rm s}$/ND & 5 & 2 & 6 & 0.58--0.74 & 1.31--1.34 & 083.C-0659\\
2009-06-01T02:08:17 & CD-40 8031              & S13 & $K_{\rm s}$/ND & 5 & 2 & 3 & 0.90       & 1.05       & 083.C-0659\\
2009-06-01T03:24:20 & V* NZ Lup               & S13 & $K_{\rm s}$/ND & 5 & 2 & 3 & 0.78       & 1.06       & 083.C-0659\\
\hline
\multicolumn{10}{c}{Keck/NIRC2}\\
\hline
\hline
2010-07-12T08:59:21 & CD-27 11535 & Narrow & $K_{\rm s}$ & 0.100 & 100 & 6 & & 1.51--1.52 & U104N2\\
2010-07-12T09:58:40 & HD 172649   & Narrow & $K_{\rm s}$ & 0.032 & 100 & 6 & & 1.05--1.06 & U104N2\\
\hline
2019-08-26T05:14:34 & CD-27 11535 & Narrow & $K^{\prime}$ & 0.050 & 100 & 22 & & 1.48 & U204\\
2019-08-26T05:20:45 & CD-27 11535 & Narrow & $H$   & 0.050 & 100 & 22 & & 1.48 & U204\\
2019-08-26T05:28:22 & CD-27 11535 & Narrow & $J$   & 0.050 & 100 & 42 & & 1.48 & U204\\
2019-08-26T05:53:51 & HD 160934   & Narrow & $K^{\prime}$ & 0.018 & 100 & 42 & & 1.33 & U204\\
2019-08-26T05:53:51 & HD 160934   & Narrow & $H$ & 0.018 & 100 & 42 & & 1.33 & U204\\
2019-08-26T06:06:23 & HD 160934   & Narrow & $J$ & 0.018 & 100 & 42 & & 1.33--1.34 & U204\\
\hline
2020-07-09T07:16:01 & CD-27 11535 & Narrow & $H$ & 0.05 & 100 & 41 & & 1.56--1.58 & D309\\
2020-07-09T07:26:26 & CD-27 11535 & Narrow & $z$ & 0.60 & 10 & 41 & & 1.53--1.55 & D309\\
2020-07-09T07:56:21 & HD 153318 & Narrow & $z$ & 0.60 & 10 & 42 & & 1.60 & D309\\
2020-07-09T08:19:32 & HD 153318 & Narrow & $H$ & 0.05 & 100 & 41 & & 1.60 & D309\\
\hline
2020-07-24T06:46:54 & CD-27 11535 & Narrow & $H$ & 0.05 & 100 & 25 & & 1.51 & U216\\
2020-07-24T06:53:56 & CD-27 11535 & Narrow & $J$ & 0.05 & 100 & 25 & & 1.50 & U216\\
2020-07-24T07:00:53 & CD-27 11535 & Narrow & $K^{\prime}$ & 0.05 & 100 & 25 & & 1.49--1.50 & U216\\
2020-07-24T07:09:30 & CD-27 11535 & Narrow & $z$ & 2.00 & 4 & 25 & & 1.48--1.49 & U216\\
2020-07-24T07:16:52 & CD-27 11535 & Narrow & $L^{\prime}$ & 0.30 & 10 & 25 & & 1.48 & U216\\
2020-07-24T07:25:54 & HD 153318 & Narrow & $L^{\prime}$ & 0.30 & 10 & 50 & & 1.60--1.61 & U216\\
2020-07-24T07:39:45 & HD 153318 & Narrow & $z$ & 2.00 & 4 & 25 & & 1.61--1.62 & U216\\
2020-07-24T07:47:25 & HD 153318 & Narrow & $K^{\prime}$ & 0.05 & 100 & 25 & & 1.62--1.63 & U216\\
2020-07-24T07:53:55 & HD 153318 & Narrow & $J$ & 0.05 & 100 & 25 & & 1.63--1.64 & U216\\
2020-07-24T08:00:26 & HD 153318 & Narrow & $H$ & 0.05 & 100 & 15 & & 1.64--1.65 & U216\\
\hline
2021-01-23T16:25:29 & CD-27 11535 & Narrow & $K^{\prime}$ & 0.3 & 50 & 15 & & 2.51--2.62 & D297\\
2021-01-23T16:16:42 & HD 153318 & Narrow & $K^{\prime}$ & 0.1 & 100 & 20 & & 2.45--2.38 & D297\\
\hline
2021-06-04T11:33:58 & CD-27 11535 & Narrow & BrG & 0.3 & 100 & 12 & & 1.52--1.54 & D335\\
2021-06-04T12:22:24 & 1RXS J195602.8-320720 & Narrow & BrG & 1.0 & 30 & 12 & & 1.68--1.70 & D335\\
\hline
\multicolumn{10}{c}{Gemini-S/GPI}\\
\hline
\hline
2018-08-15T01:05:15 & CD-27 11535 & \nodata & {\it H} & 59.6 & 1 & 19 & & 1.01--1.03 & GS-2017B-Q-22\\
2018-08-15T01:46:28 & HD 153318   & \nodata & {\it H} & 59.6 & 1 & 20 & & 1.07--1.11 & GS-2017B-Q-22\\
\hline
2019-08-05T02:14:09 & CD-27 11535 & \nodata & $K_1$ & 59.6 & 1 & 16 & & 1.04--1.06 & GS-2019B-Q-120\\
2019-08-05T02:41:50 & HD 153318   & \nodata & $K_1$ & 59.6 & 1 & 16 & & 1.09--1.13 & GS-2019B-Q-120\\
\enddata
\end{deluxetable*}

\begin{deluxetable*}{ccccccccccccc}
\tabletypesize{\scriptsize}
\tablecaption{Astrometric and photometric measurements of CD-27 11535}
\tablewidth{0pt}
\tablehead{
\colhead{UT Date} &  \colhead{Instrument} & \colhead{Filter} & \colhead{Plate Scale} & \colhead{True North} & \colhead{$\rho$} &\colhead{$\rho$ residual} & \colhead{$\theta$} & \colhead{$\theta$ residual}  & \colhead{Flux Ratio} & \colhead{Calib.} & \colhead{Data}\\
&&&(mas px$^{-1}$)&(deg)&(mas)&(mas)&(deg)&(deg)&&Ref.&Ref.}
\startdata
2006 May 27 & NaCo & IB2.27         & $27.06 \pm 0.06$ & $0.02 \pm 0.1$ & $88.45 \pm 0.78$ & $1.48$ & $282 \pm 2$ & $2.37$ & $0.87 \pm 0.05$ & a &\\
2008 May 18 & NaCo & IB2.27         & $27.08 \pm 0.03$ & $0.04 \pm 0.09$ & $91.85 \pm 0.46$ & $0.30$ & $229.4 \pm 0.4$ & $-1.35$ & $0.841 \pm 0.014$ & a & \\
2009 Jun 01 & NaCo & $K_{\rm s}$/ND & $13.22 \pm 0.02$ & $-0.17\pm0.03$  & $83.26 \pm 3.37$ & $-2.29$ & $203 \pm 2$ & $-1.37$ & $0.880 \pm 0.023$ & b &\\
2010 Jul 12 & NIRC2& $K_{\rm s}$    & $9.952 \pm 0.002$ & $0.252 \pm 0.009$ & $81.28 \pm 0.11$ & $-0.30$ & $176.30 \pm 0.11$ & $0.21$ & $0.805 \pm 0.007$ & c & \\
2018 Aug 15 & GPI  & $H$            & $14.161 \pm 0.021$    & $0.28 \pm 0.19$ & $144.23 \pm 0.31$ & $-0.19$ &  $34.26 \pm 0.22$ & $0.09$ & -- & d\\
2019 Aug 05 & GPI  & $K_1$          & $14.161 \pm 0.021$    & $0.45 \pm 0.11$ & $141.85 \pm 0.23$ & $-0.59$ &  $25.47 \pm 0.23$ & $0.16$ & -- & d &\\
2019 Aug 26 & NIRC2& $J$        & $9.971 \pm 0.005$ & $-0.26 \pm 0.02$ & $141.42 \pm 0.14$ & $-0.11$ & $24.82 \pm 0.05$ & $-0.02$ & $0.863 \pm 0.006$ & e & \\
2019 Aug 26 & NIRC2& $H$        & $9.971 \pm 0.005$ & $-0.26 \pm 0.02$ & $141.28 \pm 0.14$ & $-0.93$ & $24.86 \pm 0.04$ & $0.08$ & $0.884 \pm 0.003$ & e & \\
2019 Aug 26 & NIRC2& $K^\prime$ & $9.971 \pm 0.005$ & $-0.26 \pm 0.02$ & $142.10 \pm 0.16$ & $-0.79$ & $24.76 \pm 0.04$ & $0.04$ & $0.873 \pm 0.002$ & e & \\
2020 Jul 09 & NIRC2& $H$        & $9.971 \pm 0.005$ & $-0.26 \pm 0.02$ & $136.5 \pm 0.4$  & $-0.71$ & $16.48 \pm 0.05$ & $0.08$  & $0.871 \pm 0.008$ & e & \\
2020 Jul 09 & NIRC2& $z$        & $9.971 \pm 0.005$ & $-0.26 \pm 0.02$ & $137.0 \pm 0.4$  & $-0.16$ & $15.94 \pm 0.18$ & $-0.46$ & $0.857 \pm 0.013$ & e & \\
2020 Jul 24 & NIRC2& $H$        & $9.971 \pm 0.005$ & $-0.26 \pm 0.02$ & $136.9 \pm 0.2$  & $0.04$  & $15.94 \pm 0.06$ & $-0.05$ & $0.859 \pm 0.005$ & e & \\
2020 Jul 24 & NIRC2& $J$        & $9.971 \pm 0.005$ & $-0.26 \pm 0.02$ & $137.1 \pm 0.9$  & $0.22$  & $15.9 \pm 0.1$   & $-0.08$ & $0.84 \pm 0.02$ & e &\\
2020 Jul 24 & NIRC2& $K^\prime$ & $9.971 \pm 0.005$ & $-0.26 \pm 0.02$ & $137.7 \pm 0.7$  & $0.76$  & $15.90 \pm 0.07$ & $-0.09$ & $0.853 \pm 0.013$ & e & \\
2020 Jul 24 & NIRC2& $z$        & $9.971 \pm 0.005$ & $-0.26 \pm 0.02$ & $137.0 \pm 0.9$  & $0.10$  & $15.95 \pm 0.08$ & $-0.04$ & $0.822 \pm 0.014$ & e & \\
2020 Jul 24 & NIRC2& $L^\prime$ & $9.971 \pm 0.005$ & $-0.26 \pm 0.02$ & $136 \pm 2$      & $-1.35$ & $15.8 \pm 0.3$ & $-0.22$ & $0.81 \pm 0.04$ & e & \\
2021 Jan 22 & NIRC2& $K^\prime$ & $9.971 \pm 0.005$ & $-0.26 \pm 0.02$ & $131.5 \pm 0.9$ & $-0.49$ & $10.77 \pm 0.13$ & $-0.11$ & $0.92 \pm 0.02$ & e & \\
2021 Jun 04 & NIRC2& BrG        & $9.971 \pm 0.005$ & $-0.26 \pm 0.02$ & $128.68 \pm 0.15$ & $-0.67$ & $6.88 \pm 0.03$ & $0.03$ & $0.887 \pm 0.005$ & e &\\
\hline
\multicolumn{12}{c}{Literature Measurements}\\
\hline
2016 May 20 & SOAR & $I$ & & & $135.2 \pm 2.7$ & $0.87$ & $55.7 \pm 0.2$ & $0.38$ & & & f \\
2017 May 15 & SOAR & $I$ & & & $141.6 \pm 0.5$ & $0.30$ & $45.8 \pm 0.3$  & $0.20$ & & & f \\
2018 Mar 07 & SOAR & $I$ & & & $145.5 \pm 0.4$ & $1.44$ & $38.3 \pm 0.2$ & $0.17$ & & & g\\
2019 Jul 15 & SOAR & $I$ & & & $141.8 \pm 0.2$ & $-0.87$ & $25.9 \pm 0.1$ & $0.03$ & & & h\\
2020 Mar 13 & SOAR & $I$ & & & $141.0 \pm 0.1$ & $1.64$ & $19.4 \pm  0.08$ & $-0.16$ & & & i\\
2021 Feb 27 & SOAR & $I$ & & & $131.9 \pm 0.1$ & $0.01$ & $9.7 \pm 0.04$ & $-0.06$ & & & j\\
2022 Mar 12 & SOAR & $I$ & & & $120.1 \pm 1.0$ & $-0.90$ & $357.01 \pm 0.48$ & $-0.56$ & & & k \\
2023 Mar 06 & SOAR & $I$ & & & $108.6 \pm 1.0$ & $-0.56$ & $343.41 \pm 0.53$ & $-0.20$ & & & k\\
\enddata
\tablenotetext{a}{\cite{Ehrenreich2010}}
\tablenotetext{b}{\citet{Chauvin:2015jy}}
\tablenotetext{c}{\citet{Yelda2011}}
\tablenotetext{d}{\citet{DeRosa2020:kd}}
\tablenotetext{e}{\citet{Service2016}}
\tablenotetext{f}{\citet{Tokovinin:2018}}
\tablenotetext{g}{\citet{Tokovinin:2019}}
\tablenotetext{h}{\citet{Tokovinin:2020}}
\tablenotetext{i}{\citet{Tokovinin:2021}}
\tablenotetext{j}{\citet{Tokovinin:2022}}
\tablenotetext{k}{\citet{Priv_Comm_tokovinin}}
\label{tab:astrometry}
\end{deluxetable*}

\begin{figure*}[!b]
\label{fig:orbitcorner}
\centering
\includegraphics[scale = 0.4]{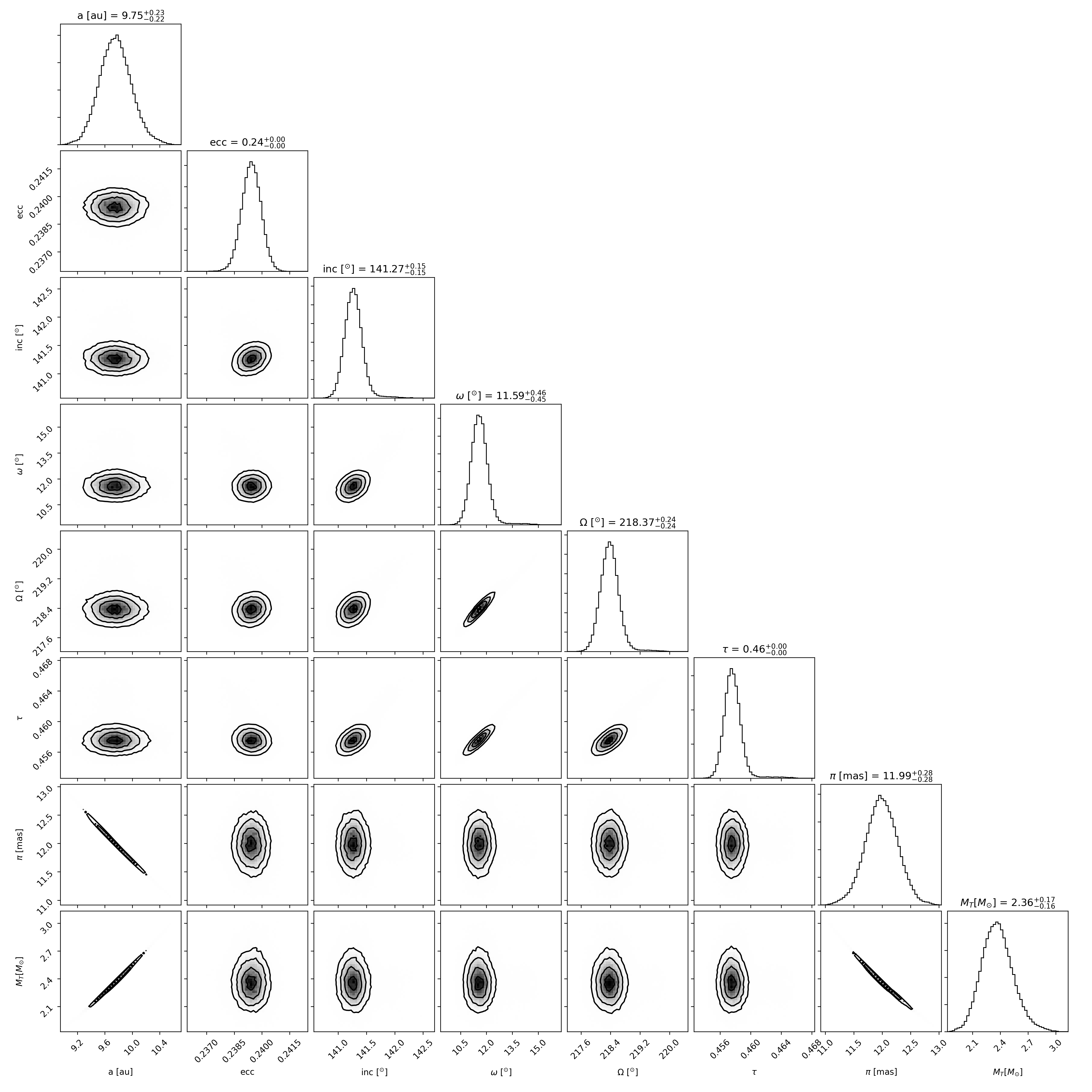}
\caption{Corner plot of orbital parameters from the \texttt{orbitize!} fit. Uniform 2D Gaussians indicate convergence.}
\end{figure*}

\begin{figure*}[!b]
\label{tab:fluxcorner}
\centering
\includegraphics[scale = 0.6]{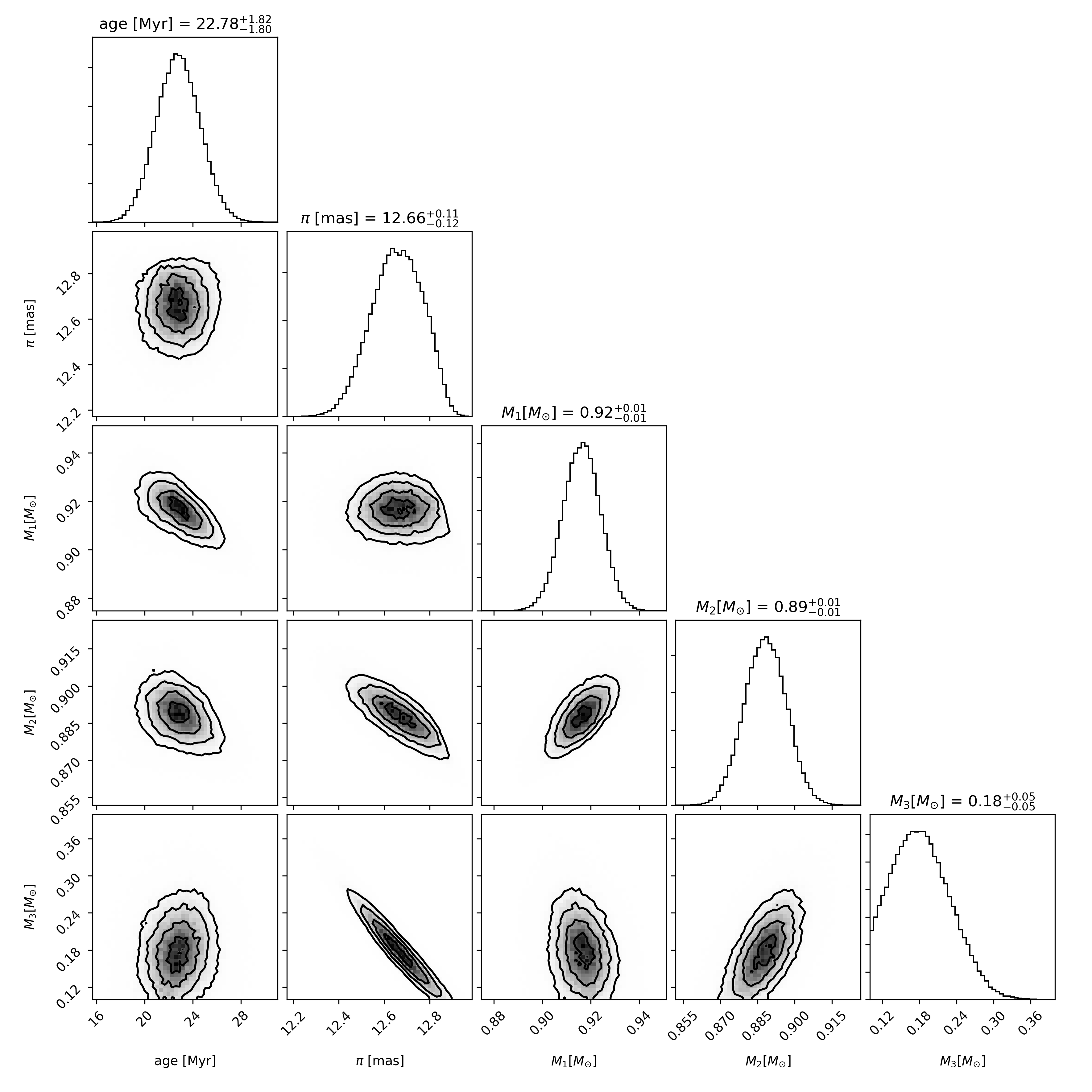}
\caption{Covariance between the system age and parallax, and the masses of the three components derived from the MCMC analysis using the Padova PAR-SEC evolutionary model with a flat prior on the parallax.}
\label{fig:cornerpadova}
\end{figure*}

\begin{figure*}[h]
\includegraphics[scale=0.45]{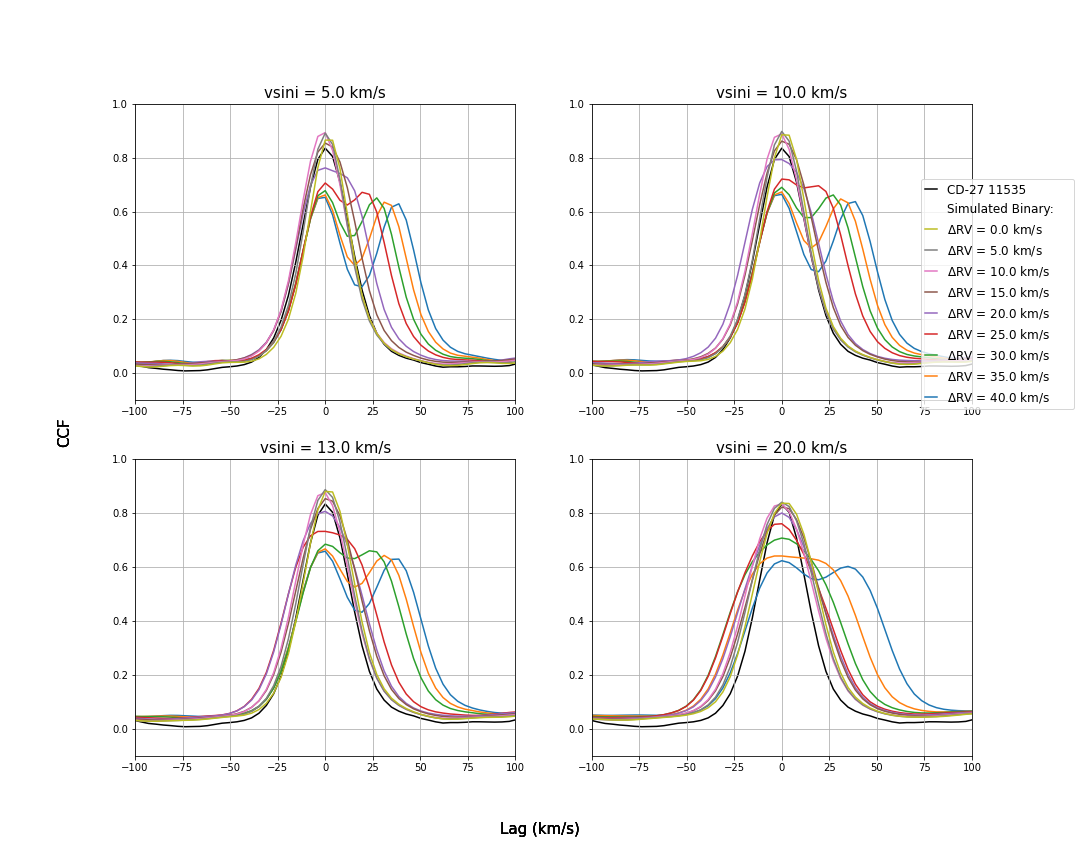}
\caption{The cross-correlation function between one observed spectrum of CD-27 11535 and the spectrum of HIP 41277, a single star with a similar spectral type, is shown in black and has a single peak. For all ARCES observations of CD-27 11535, the cross-correlation yields similar results. Cross-correlation functions between a simulated binary and HIP 41277 are shown as colored curves. The different colors represent different delta radial velocities between A and B. For each panel, we apply a different $v\sin i$ to the components of the simulated binary. Each cross-correlation function is re-centered at zero to better compare each curve. Two peaks are visible for larger delta RVs ($\gtrsim 30 km/s$) when $v\sin i = 13 km/s$. The largest delta RV that produces a single peak matching the CD-27 cross-correlation function corresponds to an orbit of Ab around Aa with a semi-major axis larger than $0.04$AU.}
\label{fig:apo_ccf}
\end{figure*}

\end{appendix}

\end{document}